# Assessing inequities in electrification via heat pumps across the U.S.

Morgan R. Edwards[1,2], Jaime Garibay-Rodriguez[1,2,3], Jacob Shimkus Erickson[2,4], Muhammad Shayan[1,6], Jing Ling Tan[1], Xingchi Shen[5,6], Yueming (Lucy) Qiu[5], Pengfei Liu[7]

## Abstract

*Heat pumps are an energy-efficient and increasingly cost-effective solution for reducing greenhouse gas emissions in the building sector. However, other clean energy technologies such as rooftop solar are less likely to be adopted in underserved communities, and thus policies incentivizing their adoption may funnel tax dollars to well-resourced communities. Unlike previously-studied technologies, the effects of heat pumps on household energy bills may be positive or negative depending on local climate, fuel availability and costs, and other factors. Here we propose a framework for assessing heat pump inequities across the U.S. We find that households in communities of color and with higher percentages of renters are less likely to use heat pumps across the board. Moreover, communities of color are least likely to use heat pumps in regions where they are most likely to reduce energy bills. Public policies must address these inequities to advance beneficial electrification and energy justice.*

# Main Paper

## 1. Introduction

Policy responses to the climate crisis are expanding rapidly in the U.S. at national, state, and local levels.[1] Research suggests that carbon dioxide ($CO_2$) emissions must decline dramatically over the next decade and reach net zero by mid-century to limit global temperature change to well below 2°C, and preferably 1.5°C, as set out in the Paris Agreement.[2] Strategies to meet these targets in the energy sector include two key elements: (1) transitioning electricity generation to low-carbon technologies such as wind and solar and (2) electrifying energy needs in transportation, industry, and buildings. Transitioning buildings off natural gas (and other fuels such as propane, heating oil, and wood) offers a meaningful pathway to reducing emissions from the building sector. Natural gas emits $CO_2$ when it is burned in home appliances and also leaks throughout the supply chain.[3,4] Unburned natural gas is primarily methane, a potent greenhouse gas.[5] A transition away from natural gas to all-electric appliances in buildings may thus be a valuable part of meeting net zero and other climate policy goals.[6]

[1] La Follette School of Public Affairs, University of Wisconsin–Madison
[2] Nelson Institute Center for Sustainability and the Global Environment, University of Wisconsin–Madison
[3] Office of Sustainability, University of Wisconsin–Madison
[4] Department of Agricultural and Applied Economics, University of Wisconsin–Madison
[5] School of Public Policy, University of Maryland at College Park
[6] Yale School of the Environment, Yale University
[7] Department of Environmental and Natural Resource Economics, College of the Environment and Life Sciences, University of Rhode Island

Heat pumps are an attractive option for electrifying space heating as they are roughly 3-5 times more efficient than traditional electric resistance heaters and can provide heating in the winter and cooling in the summer (see S1).[7] However, the net private and social benefits of heat pumps vary across locations and depend on many factors including local climate, electricity mix, prices of electricity and fuels, gas access, installation costs, and the insulation and air-tightness of a building. Heat pumps are cost effective in many locations across the U.S.[8–11] However, they can also be more expensive to operate, especially in colder regions (where they are less efficient) when compared to natural gas furnaces. High heat pump penetration will also increase electricity demand, which may place stress on electric grid capacity.[7] Moreover, the climate impacts of heat pumps (and other equipment that uses refrigerants) can increase if there is high leakage or improper management of refrigerants.[12] Nevertheless, high heat pump use rates (up to 90%) are likely to reduce greenhouse gas emissions and provide other benefits across much of the U.S., especially as the electric grid continues to decarbonize.[13]

There is a strong rationale for public policy to support heat pump adoption in the U.S. The net social benefits on a national scale are positive, yet penetration remains low (at just over 10% of households). Globally, heat pump penetration varies dramatically, with high penetration in countries such as Norway, Finland, and Sweden and low penetration in many others.[14] A rich literature on the energy efficiency gap identifies many likely barriers to heat pumps adoption,[15] including high upfront costs, credit and liquidity limitations, low familiarity with technologies,[8] low awareness of policy incentives,[15,16] personal preferences or behavioral factors,[17] and limited financial and engineering literacy.[18] Split incentives for rental units[19] or homeowners who may wish to move in the future[20] also hinder adoption (although homes with heat pumps may enjoy a price premium when sold[21]). A variety of policies have been proposed to address these barriers (see S2).[21–23] Previous incentives often take the form of tax breaks, rebates, or low- and no-interest loans at the federal, state, local, non-profit, and utility scale. Increasingly, these incentives offer additional support for low-income and qualified households, building on longstanding energy poverty programs such as the Weatherization Assistance Program (WAP) and Low Income Home Energy Assistance Program (LIHEAP).[24]

The Inflation Reduction Act (IRA) substantially expands incentives for heat pumps and other clean energy technologies at a federal level,[25] and local, state, and utility policies are also increasingly promoting household electrification. However, the extent to which these incentives will benefit underserved communities remains uncertain. Prior research indicates that low-income households are less likely to take advantage of energy efficiency and clean energy incentives.[26–30] However, the IRA provides more targeted incentives than past federal policies, and recent research on rooftop solar[31] (and prior research on income-targeted incentives for broadband[27]) suggests that income-qualified incentives may be effective at driving increased adoption among low-income households.[30,32] As heat pump sales grow rapidly in the U.S. and globally,[33] disparities in adoption may exacerbate current inequities in the energy system.[34] With higher levels of electrification, households that electrify sooner are also less likely to bear the rising costs of stranded natural gas assets.[35] An assessment of current inequities in heat pump adoption and implications for policy is thus important and timely.

Here we investigate spatial patterns in heat pump use across the U.S. to inform the design and implementation of equitable and beneficial electrification policies. Although space heating and cooling represent on average 55% of household energy use and 44% of energy bills (with high regional variation[36]), inequities in heat pump use have yet to be comprehensively explored. Two recent studies assess the social and economic determinants of heat pump adoption nationally[37] and in New York[38], but these analyses do not explicitly focus on equity. They also do not account for an important conceptual challenge: unlike technologies such as rooftop solar and energy efficiency, operating costs of heat pumps are highly variable and may exceed the costs of fossil fuel alternatives. We develop a framework for evaluating inequities in this class of technologies and apply it to heat pumps using a national property dataset and local socioeconomic data (see Methods). Specifically, we address the following questions: (1) How does heat pump use across the U.S. vary with respect to income, race/ethnicity, and renter status? (2) Where do these use patterns suggest possible inequities (and where do they not)? (3) What are the implications for policies to promote fair and just energy transitions?

## 2. Framework for estimating heat pump inequities

Energy justice calls for everyone to have access to clean, safe, affordable, and reliable energy and to be able to participate in energy decisions that affect them.[39,40] Energy justice has many dimensions: distributive, procedural, restorative, recognition and cosmopolitan.[41] Researchers and communities have identified many injustices across the energy life cycle.[40,42–44] Underserved communities are simultaneously more likely to experience energy-related pollution and less likely to benefit from energy services.[45] Energy poverty remains a important policy challenge, with over one in four U.S. households reporting difficulty paying energy bills.[24,46] Households experiencing energy poverty may limit their energy use,[47] forgo other essential services to pay energy bills, or experience shut-offs if bills are not paid.[48,49] They are also more vulnerable to shocks (e.g., COVID-19[50]), price changes,[51] and the effects of extreme weather.[52] Households in communities of color are 60% more likely to be energy burdened compared to majority white communities.[53] They are also more likely to live in energy-inefficient homes and less likely to use technologies such as rooftop solar, even controlling for income and resource potential.[15,54]

Our framework for assessing inequities in heat pump use builds on prior work on distributive energy justice by distinguishing between three concepts: inequality, inequity, and injustice[55] (see Figure 1). *Inequality* refers to a difference across populations.[56–58] Inequalities have many potential drivers and do not necessarily indicate a need for policy intervention. For example, we expect to see more heat pumps in mild climates, where they are more efficient and cost effective.[13] However, differences may persist even after controlling for these factors. Specifically, there may be fewer heat pumps in underserved households (e.g., racial or ethnic minorities, low-income, renters, etc.). Previous literature on other clean energy technologies characterizes these differences as an *inequity*.[30–32,42,59] Inequity is a normative concept that refers to what is considered unfair, with roots in philosophical and legal scholarship.[60] If a technology provides benefits, it is unfair for underserved households not to have access. Finally, an *injustice* occurs when public policies or other systemic factors exacerbate, or perpetuate, existing inequities (see Discussion for policy recommendations to advance energy justice).

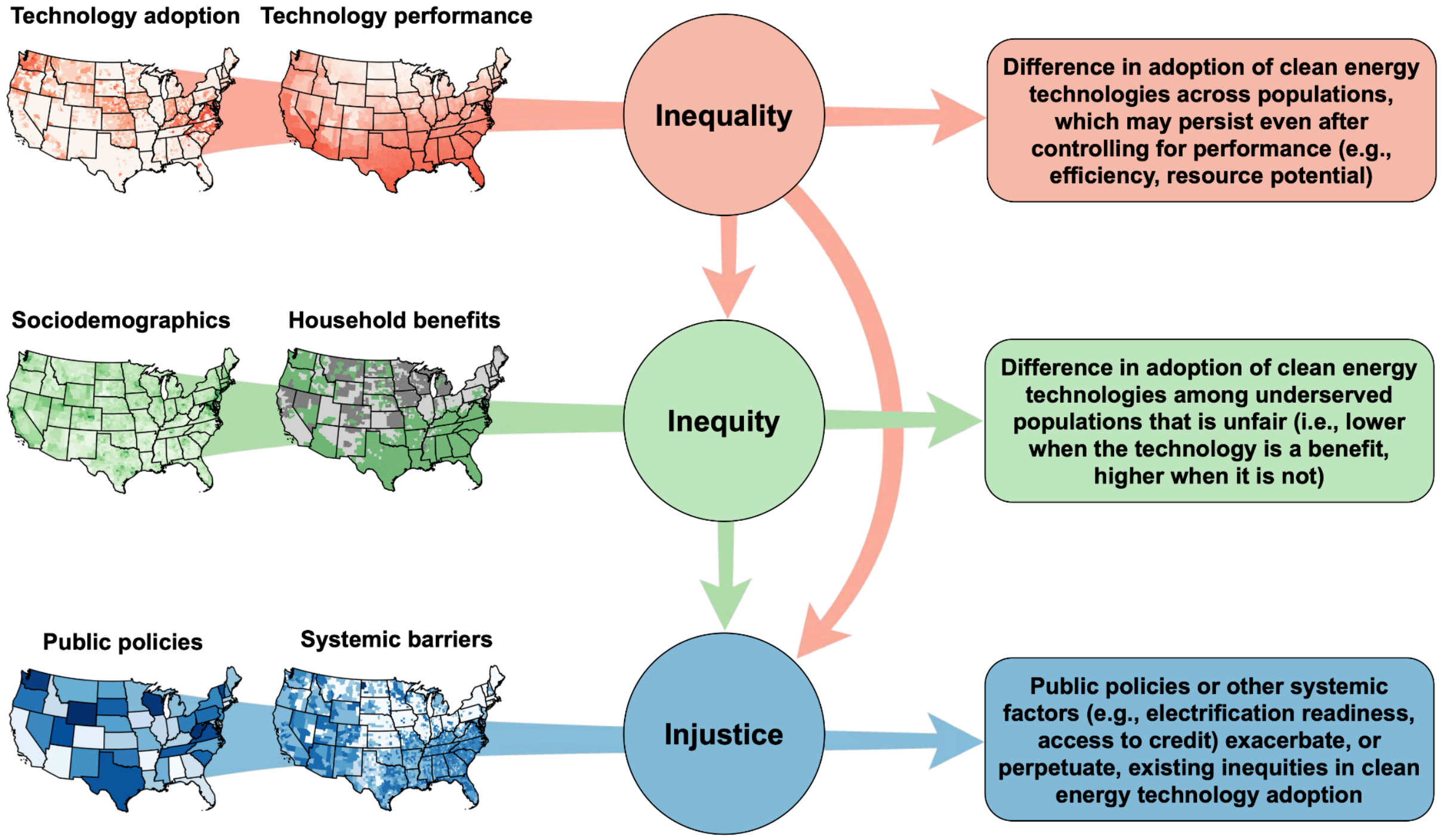

***Figure 1.*** *Framework for assessing inequalities, inequities, and injustices in clean energy technologies whose benefits to households may be positive or negative. First, we use spatial data on different technology types and performance to assess inequalities. Second, we add in data on sociodemographics and the household benefits to determine if these inequalities are also inequities. Third, this framework can be further applied to identify injustices where public policies or other systemic barriers exacerbate or perpetuate inequities. We apply this framework to assess inequalities and inequities in heat pump use across the U.S.*

Many clean energy technologies, such as energy efficiency and rooftop solar, provide benefits to households by reducing energy bills. Heat pumps, in contrast, may raise or lower bills compared to fossil fuel alternatives. Lower heat pump use among underserved households in areas where they lower heating bills (or provide other benefits such as access to cooling, for households that do not otherwise have air conditioning) represents a potential inequity. It may also suggest that there are disproportionate (and thus unjust) barriers to access, and that heat pump incentive programs unfairly divert resources away from low-income households. However, higher heat pump use among underserved households in areas where they raise bills also represents a potential inequity. For these cases, adoption by well resourced populations can provide a service to society by buying down the cost curve. To operationalize this framework, we integrate spatial data on heating technologies and a cost/benefit model of energy bill impacts. We then assess patterns in heat pump use overall and across regions where they are likely to cause energy bills to increase, decrease, or stay the same. We aim for this framework to complement and extend existing theories on justice in energy transitions.[41]

## 3. Patterns in heat pump use across the U.S.

Heating technologies vary across the U.S. (see Figure 2). As heat pump use is not reported with high granularity in other datasets, we use national household property data from the Zillow Transaction and Assessment Database (ZTRAX).[21,61] We develop a logistic regression (logit) model to estimate household heat pump use as a function of census-tract variables. Using this model, we assess whether differences in race, ethnicity, income, and renter status significantly correlate with differences in the likelihood of household heat pump use. We also control for factors expected to affect the likelihood of heat pump use such as heating and cooling degree days (HDD and CDD, which affect heat pump efficiency), electricity and natural gas prices, and building age (as a proxy for electrification readiness) and type (i.e., single family, multifamily, etc.). Finally, we include state dummy variables to account for state-level policies (such as electrification subsidies or fuel-switching restrictions) and other state-specific differences. (See Methods and S3 for a full description of data sources and S6 for robustness checks using household sample data from the Residential Energy Consumption Survey [RECS].)

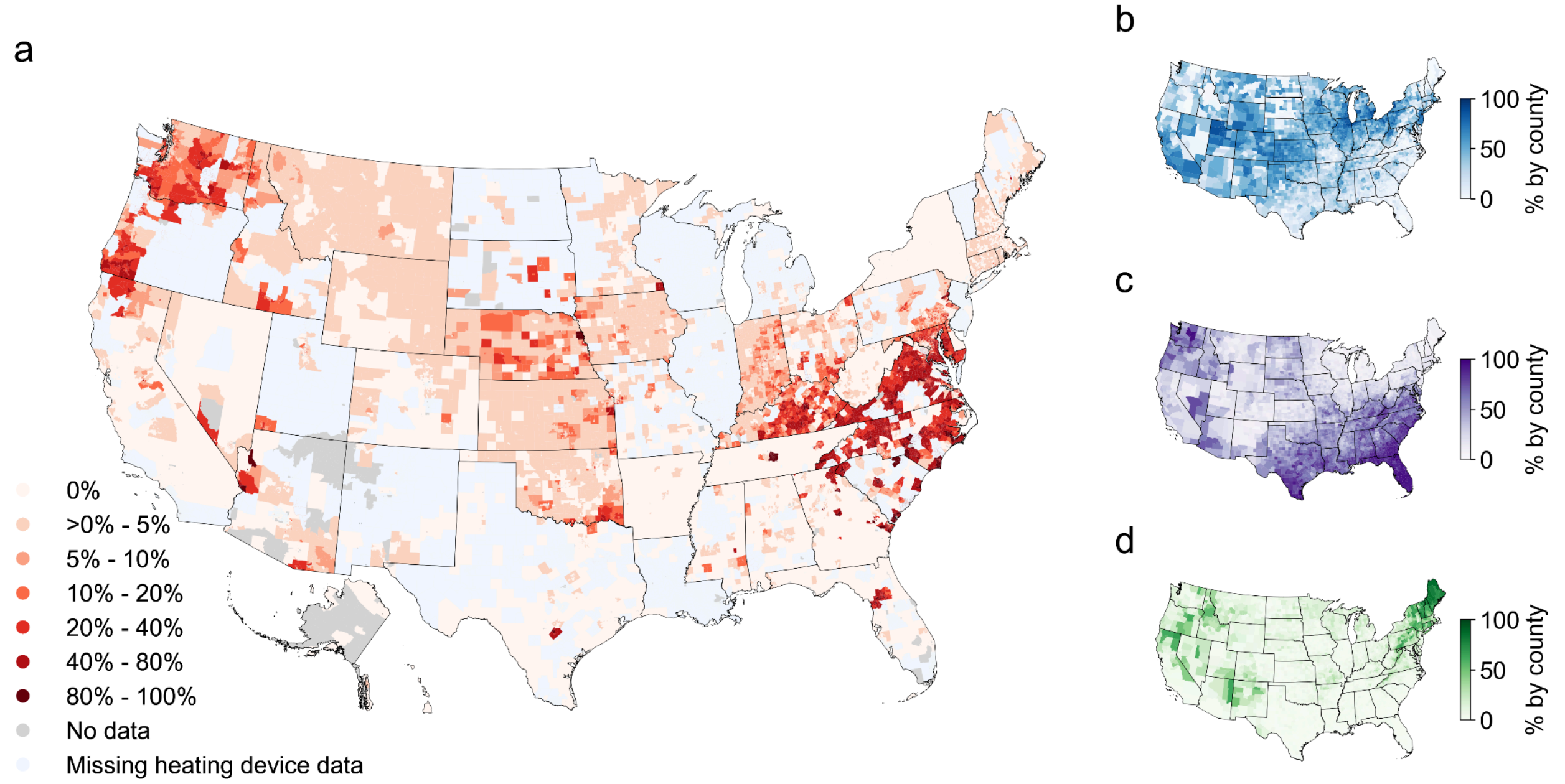

***Figure 2.*** *Patterns in heating technology use across the U.S. (a) Percentage of households with a heat pump at the census tract level from the Zillow Transaction and Assessment Database (ZTRAX) in 2020. (b-d) Household heating technology at the county level from the American Community Survey (ACS) 5-year (2016-2020) data: (b) natural gas, (c) electricity (which includes heat pumps and electric resistance heating), and (d) other fuels. (Note that some counties do not report heating technology in ZTRAX, represented by light blue shaded areas. We discuss missing data in the Methods and S3.11.)*

After controlling for state effects, several control variables correlate meaningfully with heat pump use (see Figure 3a and Table S4.1).[16] We find a negative relationship between HDD and CDD and heat pump use, with a 10% increase corresponding to a 9.4% and 2.6% decrease in odds, respectively.[8] This effect is likely explained by better heat pump performance in milder climates within states. There is also a large negative relationship with building age, with evidence of a nonlinear relationship (see S8.7). This effect is likely explained by the fact that older buildings are generally less electrification ready. Heat pump use is also lower in multifamily buildings, which face higher barriers to adoption, and higher in small dwellings such as duplexes and mobile homes. Natural gas prices and home value both have smaller correlations. Higher electricity prices are associated with higher heat pump use (with a 10% increase corresponding to a 7.3% increase in odds). This effect may be explained by areas dominated by electric resistance heating, where higher prices provide an incentive to switch to a more efficient heat pump. This explanation is supported by robustness checks, where we find that electricity price is positively correlated with heat pump use where electric resistance is dominant and negatively correlated where direct fuel use is dominant (see S8.15).

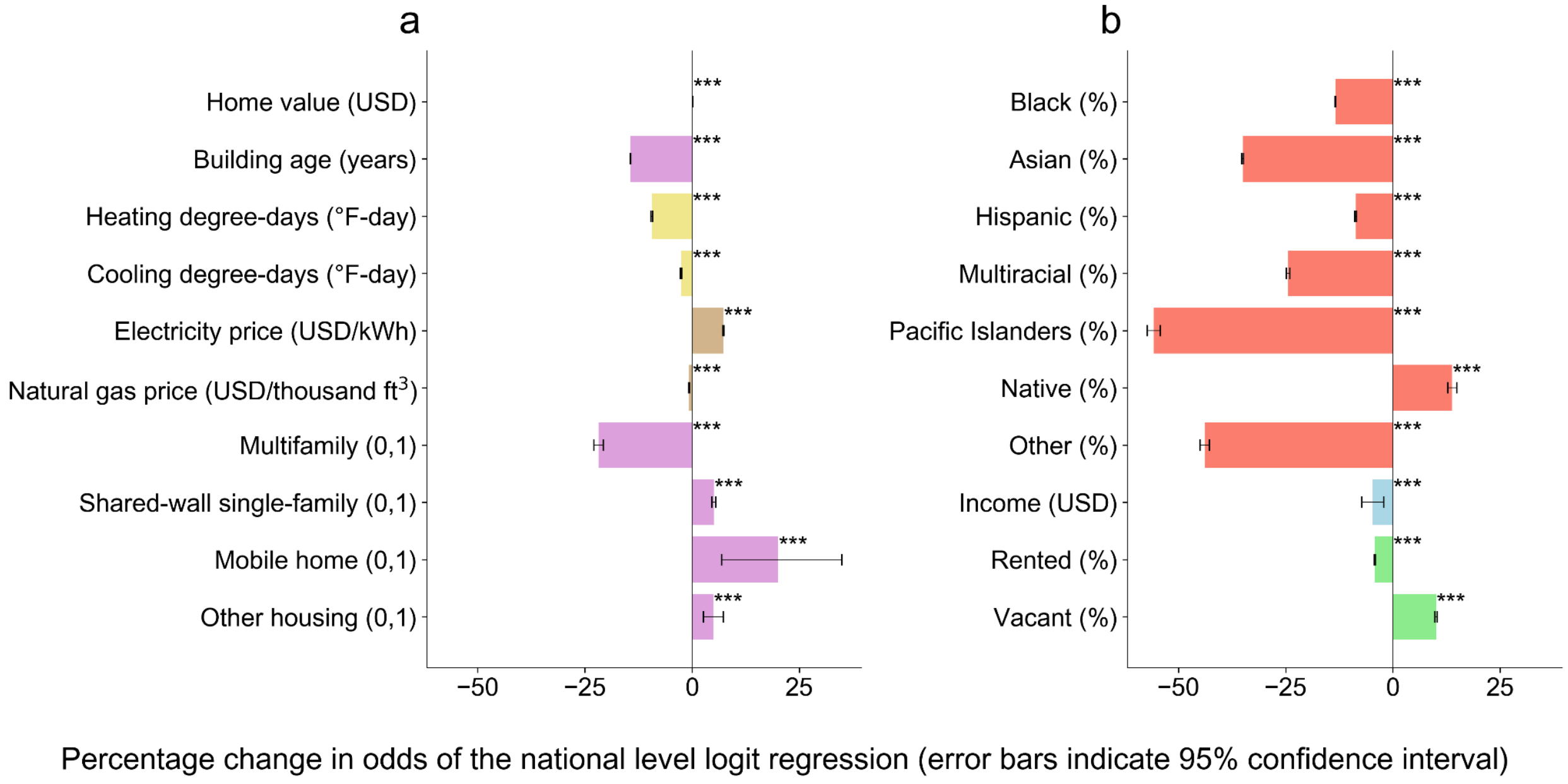

***Figure 3.*** *Percentage change in odds of heat pump use given a 10 percent increase in (a) control variables and (b) equity-related variables for the logistic regression. Descriptive statistics for each variable are provided in the S3.10. The pseudo $R^2$ value for the logistic regression is 0.3333. Symbols indicate: *** p<0.01, ** p<0.05, * p<0.1. Note that the percentage of vacant units is included in (b) to have complete representation of occupancy statuses (i.e., owner-occupied, rented, and vacant, with owner-occupied omitted due to collinearity).*

Sociodemographic variables strongly correlate with heat pump use, even after controlling for other factors (see Figure 3b). Specifically, higher minority populations within states (except for Native American), relative to white populations, significantly correlate with lower heat pump use. For example, a 10% increase in the Black population in a census tract is associated with a 13% decrease in the odds of heat pump use. The percentage of Asian, multiracial, Pacific Islander and other non-white racial/ethnic residents are also significantly and negatively associated with heat pump use. Coefficients for income and its square suggest a non-linear income effect, which we explore in Section 4 (and S10). Additionally, a 10% increase in renter-occupied units in a tract is associated with a 4% decrease in the odds of heat pump use. These findings are robust to a variety of different model formulations (see S8), and the pseudo $R^2$ value (0.3333) indicates a good overall fit .[62,63] We replicated this analysis using RECS and found largely consistent results with respect to race/ethnicity and income but not renter status (see S6).

[8] Note that all values for changes in odds are presented as a percentage point change.

## 4. Nonlinear estimates of heat pump use

We next explore non-linear relationships between heat pump use, income, and renter status across different racial/ethnic majority census tracts (see Methods and Figure 4). We normalize heat pump use by state to control for state-level variation in climate, policies, electricity and fuel prices, and other variables. Households of color are disproportionately low-income and more likely to be renters, but distinct patterns emerge when comparing areas with similar median income and homeownership rates.[64,65] For white-majority census tracts (69% of tracts), heat pump use increases slightly with median income up to approximately $60,000 and then declines. In contrast, Black-majority census tracts (7.3% of tracts) show a sharp rise in heat pump use at approximately the same income level. Hispanic and no-majority census tracts have installed fewer heat pumps compared to their white counterparts across all income levels. Similarly, for white-majority and Black-majority census tracts, heat pump use generally decreases as the percentage of rented households increases, while other racial majority census tracts have lower heat pump use overall and flatter trends. We also find evidence of nonlinear relationships between heat pump use, electricity prices, and home age (see S9).

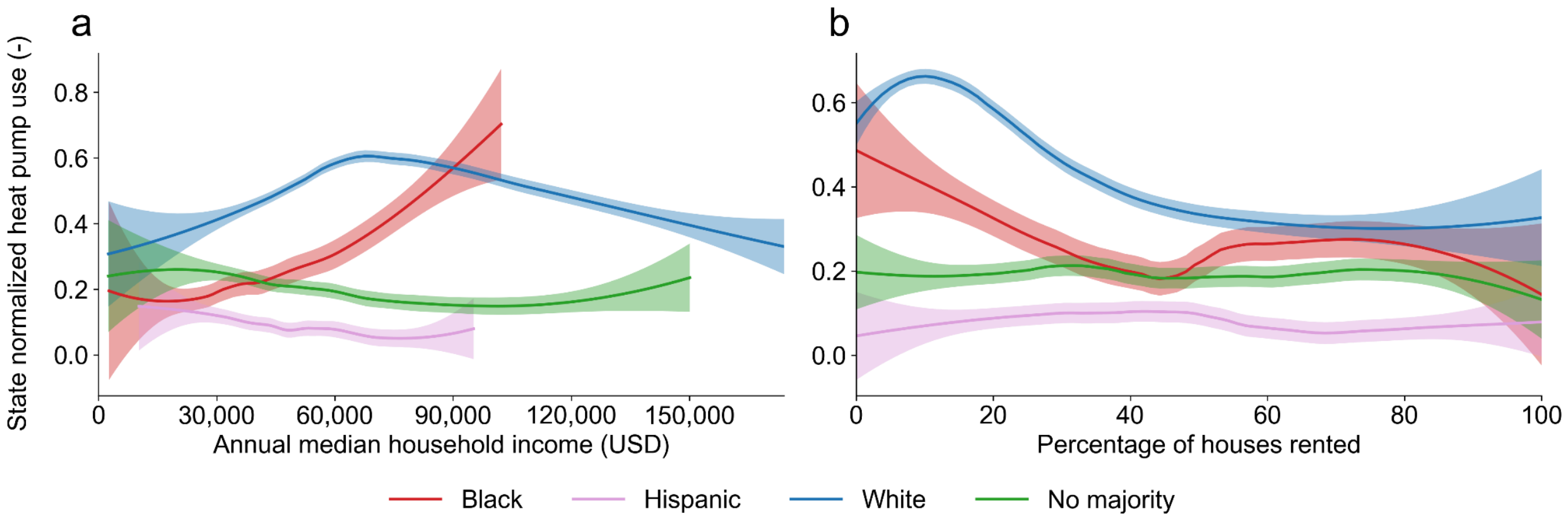

*__Figure 4.__ Nonlinear relationships between heat pump use, income, and renter status across racial/ethnic groups. Relationship between (a) median annual household income (b) percentage of houses rented and state-normalized heat pump installations by racial and ethnic majority census tract. Lines show LOESS curves and shaded regions show 90% confidence intervals (see Methods for further details).*

## 5. Regional analysis of heat pump inequities

We examine patterns in heat pump use in areas where they are highly likely to provide net benefits (or costs) to households in the form of changes to energy bills. First, we calculate the net effects of heat pump adoption on energy bills at the county scale by comparing the operating costs of a heat pump to those of the most common primary heating technology, taking into account device efficiencies, energy prices, and climate data (see Methods). Following previous studies,[66–68] we focus on differences in operating costs because (1) we are assessing inequities in heat pump use rather than new adoption and (2) differences between average installation costs of a heat pump compared to other heating devices (e.g., natural gas, propane, or fuel oil furnace or electric resistance heater) plus air conditioning are small (see S7.6). Similarly, we do not include operating costs of cooling because differences across cooling technologies (e.g., heat pump versus a stand-alone air conditioner) are smaller than those across heating technologies. We recognize that not all homes with heating also have cooling; for these homes, switching to a heat pump may entail higher energy bills but greater access to a service (i.e., cooling) that may be increasingly important under climate change.[69]

We perform a Monte Carlo analysis to identify counties that exhibit either net benefits or costs in 90% of simulations and a neutral region where at least 10% of simulations yield net benefits and net costs. Net benefits vary across the U.S., with 53% of households in our dataset seeing benefits in terms of reduced energy bills, 19% seeing costs, and 28% seeing no clear benefits or costs (see Figure 5a). While our dataset does not contain all U.S. homes (see S3.11), our results are consistent with findings from recent studies that focus on operating[66–68] or life cycle[11,13] (including installation) costs. Primary heating fuel is a key driver of benefits: in regions where the heating source is electricity or fuels other than natural gas, heating is generally more expensive and thus the net benefits of heat pumps are higher (see Figure 5b). However, climate and energy prices also play an important role. To examine inequities in heat pump use, we then run separate logistic regressions for these three cost/benefit regions (see Figure 5c-e). Using the framework presented in Section 2, we define inequities as occurring where there is (1) a lower odds of heat pump use in underserved communities in the net benefits region or (2) a lower odds in the net costs region. We use the same variables as in our national analysis (see S5) but focus our discussion on results related to race, income, and renter status.

Race and ethnicity are significantly correlated with lower heat pump use (within states) across all three regions. However, the negative relationship between minority populations and heat pump use is larger when heat pumps provide net benefits to households relative to regions with net costs. For example, a 10% increase in the Black population in a census tract leads to a 13% decrease in the odds of heat pump use in regions with net benefits (coincidentally the same value as in our national analysis), compared to a 5.1% decrease in regions with net costs. For the neutral region, we observe a larger decrease (25.8%) for the Black population; however, the overall effect across racial/ethnic minorities is less pronounced in comparison to regions with net benefits. The percentage of the population living in regions where heat pumps provide net benefits versus net costs varies across racial and ethnic groups, with 39% of the white population and 52% of the Black population living in regions where heat pumps provide net benefits, compared to 39% overall (see SI Tables 5.2 and 5.3). However, all racial and ethnic groups are represented in each of our three regions (see S5 for further discussion).

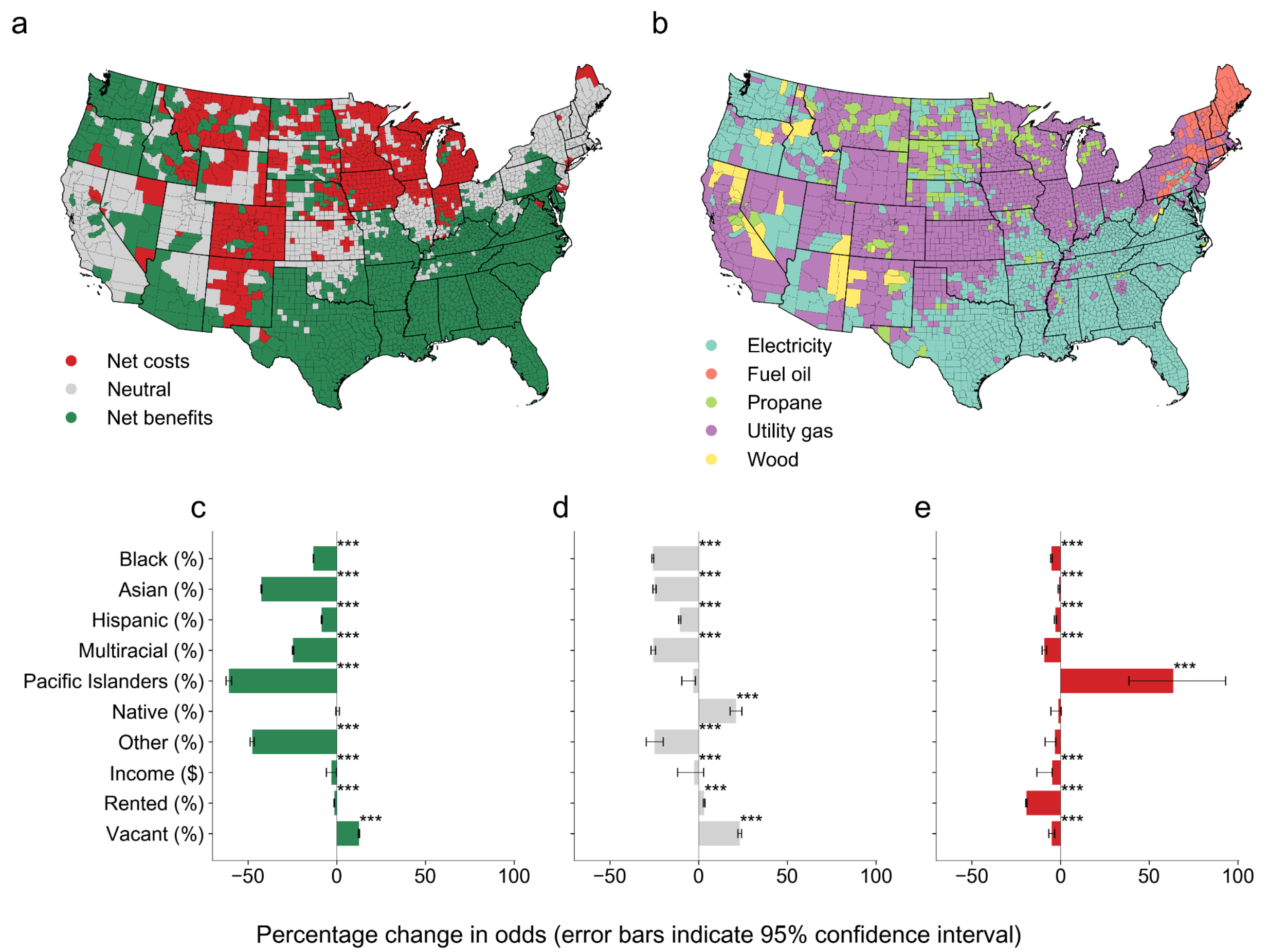

***Figure 5.*** *Inequities in heat pump use across the U.S. (a) Counties where heat pump use is a net benefit, neutral, or net cost for households in terms of changes in energy bills and (b) primary heating source. Percentage change in odds for logistic regression results for regions where heat pumps are a (c) net benefit, (d) neutral, and (e) net cost. The Pseudo $R^2$ values for the logistic regressions are 0.3189, 0.1953, and 0.2642 for the net benefits, neutral, and net costs regions, respectively. Symbols indicate: *** p<0.01, ** p<0.05, * p<0.1. No symbol indicates p>0.1.*

We also observe differences in the relationship between income and renter status and heat pump use across the three cost/benefit regions. To explore these differences further and assess potential nonlinearities, we repeat the LOESS analysis described in Section 4 for each of the three cost/benefit regions (see Figure S9.1). We observe a generally positive relationship between income and heat pump use across all three regions up to a median income of approximately $100,000 (for net benefit regions), $65,000 (for neutral regions), and $77,000 (for net cost regions). After these levels, the effect of income is negative but smaller for the net benefit and neutral regions and larger for the net cost regions. Renter status negatively correlates with heat pump use in the net benefits and net costs regions but has the largest coefficient in the latter, whereas it exhibits a small positive relationship in the neutral region. As in the results on race and ethnicity, these results also suggest that low-income and renter

communities are less likely to use heat pumps, even in regions where they provide benefits, suggesting an equity that will be important to address in electrification policies.

## 6. Discussion

Household electrification is an emerging energy justice issue in the U.S. At the national scale and controlling for state fixed effects, we find that higher percentages of racial and ethnic minority households, lower median income, and higher percentages of renters at the census tract level all correlate with lower heat pump use. The negative correlations between racial and ethnic minority populations and heat pump use are largest in areas where heat pumps are expected to provide direct benefits to households. While directionally the same, the negative correlation between renter status and heat pump use is largest in areas where heat pumps are likely to provide a net cost to households. Income effects are nonlinear but suggest that moderate income is generally associated with higher heat pump use across net cost and benefit regions. Taken together, these results suggest that, within states, minority communities are disproportionately less likely to reap the benefits of using a heat pump. Additionally, our nonparametric analysis shows that higher income among Black-majority census tracts is associated with a higher likelihood of heat pump use relative to other groups, suggesting distinct relationships between heat pump use, race/ethnicity, and income.

Our analysis contributes new insight to the field of energy justice by assessing inequities in a technology that may have higher or lower operational costs than incumbents. Widely studied technologies such as rooftop solar or energy efficiency may have high upfront costs but have lower operational costs across the board. By separately analyzing dynamics in regions where heat pumps have positive, negative, or neutral effects on energy bills, we can provide a more nuanced assessment of whether lower (or higher) heat pump use in underserved communities suggest potential inequities. Because we find evidence of racial/ethnic, income, and (to a lesser extent) renter status inequities in regions where heat pumps provide bill savings, policies, markets, or other drivers may be implicitly under-serving these marginalized groups.[32,57,70,71] Conversely, lower heat pump use in regions where heat pumps raise energy bills does not represent an inequity and can help drive down costs, particularly for heat pumps designed to operate in cold climates. Notably, race and ethnicity are the strongest negative correlates of heat pump use in regions where heat pumps are expected to result in energy bill savings, leading us to conclude that there is evidence of an emerging inequity.

We present a quantitative assessment of national racial and ethnic patterns in heat pump use and find robust evidence of heat pump inequities for racial/ethnic minorities, with smaller effects for low-income households and renters. Looking forward, we note several areas for future research. Because we do not measure demographic variables at the household level, our results may under- or over-estimate inequities. Robustness checks using nationally representative household data from RECS along with prior literature (which shows that environmental inequities often increase with data granularity) provide further evidence for the inequities we observe.[64,65] However, future work may leverage more granular data to assess these dynamics at the household level. We also focus on current heat pump users, but the demographics of new adopters may be different. Future work could explore these dynamics and investigate patterns of technology diffusion over time and across different countries.[72] It can also investigate the effects of changes in the relative prices of electricity and fuels over time, either due to market forces or tax policies, and the effects of technology adoption on heating and cooling demand.[73] Finally, while the mechanism behind the correlations we observe is outside the scope of this paper, the evidence presented warrants further causal analysis.

Our framework can be applied to other clean energy technologies where the net household benefits may be positive or negative.[74,75] While we focus specifically on effects on energy bills, there are a variety of other potential costs and benefits of new technologies. For example, heat pumps may increase cost reliability (particularly in regions where fuel prices are highly variable[14]), improve thermal comfort, and provide access to cooling for households without air conditioning. They may also have other costs such as increased backup requirements during electricity service interruptions. Utility incentive programs can also affect adoption decisions, and future work could provide targeted recommendations for state and utility policy to advance beneficial electrification. Beyond the individual household, heat pumps also have costs and benefits to society, including decreasing or increasing air pollution and greenhouse gas emissions, depending on the electricity source. As more households electrify, cost dynamics can change and potentially create larger injustices. For example, as the customer base for natural gas utilities shrinks, the burden of cost recovery for durable infrastructure is placed on a smaller number of remaining customers and may raise energy bills.[76–78] Underserved communities may disproportionately bear these stranded asset costs.

Addressing heat pump inequities is important globally and perhaps particularly in the U.S., given recent policies.[79] Our analysis suggests that care will need to be taken to ensure that the $43 billion in consumer incentives for energy efficiency and electrification directly benefits underserved communities. By using insights on current patterns in heat pump use and their costs and benefits, policies can be designed to reduce inequities, and ultimately injustices, in electrification transitions. We identify three key policy recommendations:

1. Building electrification policies must be location and context specific. For locations where the heat pumps reduce energy bills, targeted outreach can be coupled with free or low-cost installations.[19] However, in areas where they may increase bills, incentives must be accompanied by efforts to reduce energy costs, such as energy efficiency upgrades and on-bill financing. It is important that the effects of heat pump adoption on energy bills are accurately communicated,[80] both to avoid increasing energy poverty and to prevent community backlash, as has been observed with time-of-use pricing programs.[81,82] The structure of the IRA, which includes funding for state and local governments, may help initiatives be more responsive to community needs and priorities.
2. Targeted incentives can increase adoption in underserved households. Historically, uptake of incentives and outreach by private firms has focused on higher income households who are more likely to be early adopters.[83] By deliberately reversing this pattern with targeted incentives and outreach, the IRA may more effectively reach otherwise underserved populations. Research suggests that means-tested incentives can be effective at increasing adoption of other technologies (e.g., energy efficiency and rooftop solar) among targeted populations.[83] Targeted outreach to contractors, installers, and landlords will also be important to promote access for underserved households.

3. Beneficial electrification requires a community and system perspective. Heat pump adoption can have important interactions with other energy policies and infrastructure. For example, areas without access to natural gas tend to have higher heating costs and can be targets for early electrification, even in relatively cold climates. Additionally, prior research has noted that policies such as increasing block pricing in California create incentives to adopt rooftop solar for specific demographics.[84] Clean energy technologies such as rooftop solar have also been shown to follow a seeding pattern,[72] and thus these policies could encourage future adoption. These same perspectives will be important to consider in future research and policy discussions.

# Methods

## 7. Data sources

We analyze spatial patterns in heat pump use across the U.S. using household-level property data from the Zillow Transaction and Assessment Database (ZTRAX), accessed January 2021.[85] The dataset includes approximately 163 million buildings (147 million residential) across the U.S., 63.7 million of which report at least one of 38 heating and cooling technologies (e.g., heat pumps, furnaces, boilers, radiators, space heaters, central A/C units, window A/C units, etc.). We designate a household as using a heat pump if it is listed as the primary heating source. ZTRAX also provides information on property characteristics including building type (single-family, multifamily, shared-wall single-family, mobile home, other), location, and market and assessed value. Additional characteristics such as number of stories, number of rooms, number of units, building area code, building area square feet, occupancy status, effective year built, and value certification date were excluded from the analysis as over 15% of the data in these variables were missing. For residences that did not have data on home value, the assessed value was used as a substitute. The ZTRAX database includes hundreds of additional variables including home sale features and information on architecture and finishes that were deemed less relevant to this analysis. See S3 for a full description of variables.

We then added census tract data (racial and ethnic identity, median income, homeownership status, and house age) from the American Community Survey (ACS) 5-year estimates for 2016–2020[86] via the R tidycensus package version 1.2.3.[87] The ACS data was merged into the ZTRAX data using the household 2020 census identifiers (GEOIDs) calculated using latitude and longitude. This merge resulted in a data loss of 0.1% of ZTRAX records due to invalid location data. We also included heating and cooling degree days (HDD and CDD) calculated from average daily outdoor temperature data from the nearest NOAA weather station between 1991-2020[88] (number of days with an average temperature above or below the standard 65°F[66] outdoor temperature benchmark). We obtained natural gas utility boundaries from the Homeland Infrastructure Foundation-Level Database (HIFLD)[89] and price data (for 2020) from the EIA-176[90] form. Average prices for residential electricity customers at the utility level were extracted from EIA Form 861.[91] Prices were normalized to 2020 using the EIA national electricity price index.[92] To calculate prices by county, we weighted the electricity price for each utility by the number of customers served. After data cleaning and merging, our final dataset includes 99.7% of all residences in ZTRAX that report heating type.

## 8. Parametric analysis

We used a logistic regression (logit) model to assess how the log odds of a residence having a heat pump relates to the racial and ethnic makeup, median income, and percentage of renters in that census tract as well as other control variables at the house, census tract, and county scale. Logit models are used to identify the probability of an event, such as having a heat pump, and are commonly used in research on household use of energy-efficient and clean energy technologies.[93] The general logit model takes the form:

$$P(H) = \frac{1}{1+e^{-Y}}, \tag{1}$$

where $P(H)$ is the probability of heat pump installation at the household level and $e$ is the Euler's number. Y is a linear function of variables as follows:

$$Y = \beta_0 + \sum_{i=1}^{N} \beta_i R_i + \sum_{j=1}^{M} \beta_j C_j + \varepsilon, \tag{2}$$

where $\beta_0$ is the intercept, $\beta_i$ are coefficients for equity-related variables $R_i$, $\beta_j$ are coefficients for control variables $R_j$, and ε is the error term. We present results in terms of percentage change in odds for a 10% increase in all continuous variables (see Table 4.1 and S4 for coefficients).

The equity related variables $R_i$ include: the fraction of households of each racial/ethnic identity (with white omitted due to collinearity), median income (log transformed), median income (log transformed) squared, and the percentage of homes that are rented and vacant (with owned homes omitted due to collinearity) at the census tract level. Racial/ethnic identity categories include (1) Black, (2) Asian, (3) Hispanic, (4) Multiracial, (5) Pacific Islander, (6) Native Hawaiian or Other Pacific Islander, (7) American Indian or Alaska Native, and (8) other (i.e., households identifying themselves as a race or ethnicity different from the categories as mentioned above). The control variables $C_j$ include housing price, building type, and building age (at the household level); county-level data on electricity costs; utility-level data on natural gas costs and access; and state fixed effects. See S3 for more details on these variables. The logit models are implemented in STATA version 17. We also developed additional models (see S8) including interaction effects between race/ethnicity, income, renter status, type, and building age.

## 9. Nonparametric Analysis

We used nonparametric local smoothing models to illustrate how heat pump use varies across median income and homeownership status for census tracts with different racial/ethnic majorities. We used LOESS (locally weighted least squares) regression to fit local linear relationships between state-normalized heat pump use and (a) median household income, (b) the percentage of houses occupied by renters, and (c) building age (at the household level) for census tracts with majority populations of different racial and ethnic groups. We include racial and ethnic groups that have a majority in at least 1% of census tracts (i.e., white, Black, and Hispanic; see SI Figure 9.1). Because LOESS smoothing does not require a global function to fit all data, it allows us to observe the otherwise difficult-to-interpret non-linear relationships between income or homeownership and heat pump use.[15] LOESS was implemented using the R stats package version 4.2.1.[94] For better visualization, outliers were removed by limiting the annual median income to the bottom 98 percentile for each racial/ethnic category.

Following similar research focused on rooftop solar,[15] localized effects such as heating and cooling demand, electricity prices, and public policies are partially mitigated by calculating the state-normalized heat pump use ($S_i$), which normalizes census-tract-level heat pump use by the population *(P)* weighted heat pump use in each state:

$$S_i = HP_i / \sum_{i=1}^{T} \left( \frac{P_i}{P_s} * HP_i \right), \; i = 1,\dots, N \;, \quad (2)$$

where $HP_i$ is the heat pump use percentage in census tract *i*, *s* is the state, and *T* are the total number of census tracts in the state. A value greater than one indicates that the census tract has installed more heat pumps compared to the state average (and vice versa). We applied this method to calculate separate nonparametric curves for census tracts based on their racial/ethnic majority. For example, if more than 50% of the population of a census tract self-identified as Black, the census tract was designated as a Black majority tract. We also included census tracts with no racial/ethnic majority.

## 10. Robustness checks

To account for the limitations in the ZTRAX data, we test for robustness using different model specifications and using RECS as an alternative, nationally representative dataset. We first replicated our parametric analysis using data from RECS.[95] RECS is a periodic survey conducted by the U.S. Energy Information Administration (EIA) that includes household-level data on geography, climate, income, household quality, and renter/owner status. The RECS sample is designed to meet precision requirements on energy consumption for all 50 states and Washington, DC, with a total of 18,496 respondents.[96] We use this data to assess how heat pump use correlates with equity (racial/ethnic makeup, income, and renter status) and control variables at the household level. The logit model is implemented in R using the survey package,[97] which is recommended by EIA to capture the sampling weights used in the RECS survey methodology.[98] The results are generally similar to our main analysis (see S6) and other analyses focused on the effects of different correlates such as income.[37,38]

We also conduct robustness checks using the ZTRAX data with different model specifications:

- First, run an analysis where we sequentially add variables to our analysis. We begin with a basic national model that includes race/ethnicity, climate, energy prices, and state dummies (S8.1) and sequentially add income (S8.2), building age (S8.3), tenancy status (S8.4), home value (S8.5), and housing type (S8.6).

- Second, we also run an additional analysis that replaces the continuous building age variable with a categorical variable for buildings built each decade from pre-1940s to the 2000s from the American Community Survey (S8.7).
- Third, we run an analysis that replaces natural gas price with a dummy variable for natural gas access (S8.8).
- Fourth, we run analyses that consider interaction terms for race/ethnicity and income (S8.9), race/ethnicity and tenancy status (S8.10), race/ethnicity and building age (S8.11), and race/ethnicity and income, tenancy status, and building age simultaneously (S8.12).
- Fifth, we run an analysis that replicates the model in our main paper but does not include state dummy variables (S8.13). We also run a version of this analysis that considers state-level rather than county-level electricity prices (S8.14).
- Sixth, we run separate analyses for regions where electricity or other fuels (including natural gas) are dominant with (8.15) and without (8.16) including energy prices.
- Finally, we run analyses using more granular data at the census block group level instead of the more aggregated census tract level, for both our national (8.17) and regional (8.18) results.

We also use the Holm–Bonferroni method to address Type I errors (false positives) given the large number of predictors for our national and regional regressions. We use a family-wise error rate (alpha) equal to 0.05.

## 11. Net benefits analysis

We estimate the net benefits of switching to a heat pump as the change in operating costs (i.e., energy bills). Our analysis uses inputs on the primary heating technology (from the ACS 2016-2020 survey) and average annual electricity prices at the county level. We use state-level data on natural gas, propane, fuel oil, and wood prices. All prices are from 2020. We calculate the coefficient of performance (COP) of a heat pump in each county using hourly temperature profiles from typical meteorological year data.[99] Following previous studies, we use a function that assumes a linear relationship between temperature and COP for temperatures between 5 and 47°F for heat pumps of similar capacity (25,000 BTU) based on the Northeast Energy Efficiency Program (NEEP) (see S7.3).[13,66] We assume average efficiencies of 80% for natural gas and propane furnaces, 83% fuel oil furnaces, and 79% for wood stoves (based on the 2015 Federal standard for residential furnaces[100] and average efficiency reported by the EPA-certified wood stove database).[101] We exclude differences in installation costs but find these to be small when comparing a heat pump with a conventional heating and cooling system (see S7.6). We also assume constant heating demand and electricity prices.

Using these inputs, we calculate the energy bill (electricity and fuel) savings or costs for a typical[36] single-family, detached house:

$$s = \sum_{i=1}^{8760} \frac{Pc \cdot q_i}{\eta_c} - \frac{Pe \cdot q_i}{COP(T_i)}$$

where $s$ are the net energy bill savings (or, if negative, costs), $Pc$ and $Pe$ are the prices of current heating fuel and electricity, $q_i$ is the hourly heating load, $\eta_c$ is the efficiency of the current heating device (e.g., natural gas furnace), and $COP(T_i)$ is the heat pump efficiency, which depends on outside temperature at hour *i*. While the absolute costs can depend strongly on heating load, which is a function of house characteristics (e.g., level of insulation, age, etc.), the relative savings or costs of different heating sources are likely to be similar. Our approach does not include differences in housing characteristics, but we do include a higher level of spatial granularity compared to previous studies on the costs of heat pump use.[13,66,68] See S7 for complete details on our approach. Future research on heat pump equity could calculate net benefits in more detail by reflecting, for example, differences in installation costs, building characteristics, and heating demand across technologies.[100]

To account for uncertainties in heating device efficiencies and fuel prices, we use a Monte Carlo simulation. For heating technology efficiencies, we construct triangular distributions using 78% as the minimum efficiency for natural gas, propane and fuel oil furnaces (based on the 1992 Federal standard[100]) and 58% for wood stoves (the minimum value reported in the EPA-certified wood stove database[101]). For maximum values, we use 99% for natural gas and propane, 96.7% for fuel oil furnaces (based on the highest efficiency in the Energy Star certified furnaces database[102]) and 90% for wood stoves (based on the maximum efficiency in the EPA-certified wood stove database). We use a triangular distribution with a minimum and maximum COP of 3 and 5 at 47°F for the heat pump COP, based on values in the NEEP database (note that the minimum COP can be as low as 1 for air temperatures below 47°F). For the price of electricity, natural gas and other delivered fuels, we use triangular distributions based on the historical national variation (the coefficient of variation using two standard deviations and the mean of national average prices as a percentage of 2020 prices) from 2000 to 2020. See S.5 for full details on the uncertainty distributions. We run the Monte Carlo simulation 10,000 times.

## Data and code availability

Individual property data were provided by Zillow through the Zillow Transaction and Assessment Dataset (ZTRAX). More information on accessing the data can be found at http://www.zillow.com/ztrax. All other data sources are publicly available online. The results and opinions are those of the author(s) and do not reflect the position of Zillow Group. The code and publicly available data to reproduce the analysis of this paper will be made available following publication.

## Funding acknowledgements

The authors acknowledge support from the Office of Sustainability at the University of Wisconsin–Madison (JGR), a University of Wisconsin–Madison University Fellowship (JSE), a generous gift from Wes and Anke Foell (JSE and JLT), the Herb Kohl Public Service Research Competition (MRE and JLT), the Alfred P. Sloan Foundation (XS, YQ, and PL), and the Wellcome Trust (MRE and JSE).

## Conflict of interest

The authors declare no conflicts of interest.

# Supplementary Information

## Assessing inequities in electrification via heat pumps across the U.S.

Morgan R. Edwards, Jaime Garibay-Rodriguez, Jacob Shimkus Erickson, Muhammad Shayan, Jing Ling Tan, Xingchi Shen, Yueming (Lucy) Qiu, Pengfei Liu

### Contents

## Section 1: Heat pump technologies and costs

Heat pumps are an efficient, electric heating and cooling technology. They are generally substantially more efficient than other sources of heating and cooling and can be a lower-emissions alternative to traditional heating systems, particularly when powered by electricity from low-carbon sources.[1,2] While traditional furnaces are powered by the combustion of fuels such as natural gas (or use electric resistance heating), heat pumps use electricity to transfer heat. The underlying operation of a heat pump is based on using mechanical work (i.e., a compressor unit) to increase the pressure (and thus, the temperature) of a refrigerant fluid and circulate it through a heat exchanger inside a building to provide space heating.[3] Typically, heat pumps can also be used for indoor space cooling by reversing the direction of the refrigerant fluid, which draws heat from indoors and expels it to the ambient environment. The ability of heat pumps to provide both heating and cooling is a key advantage over traditional heating and cooling systems that rely on two separate devices (e.g., a furnace and an air conditioner) and can be particularly beneficial for homes that do not already have air conditioning systems. This analysis, however, focuses on heat pumps used as the primary source of heat as this represents the greatest opportunity for displacing on-site fossil fuel combustion.

Three types of heat pumps are currently available on the market: air-, ground-, and water-source. Air-source heat pumps (ASHPs), which transfer heat between indoor and outdoor air, are the most common type used in the U.S[2] and have a global market share of over 60%. ASHPs may be connected by ducts but are also available as ductless devices (i.e., mini-split) for homes without ducts. While there is heat in outdoor air even in winter, ASHPs become markedly less efficient in extremely cold climates.[4] Traditionally, dual fuel systems are used in these cases, where a gas furnace or electric resistance heating system provides backup heat. However, improvements in two-stage heat pump systems, which have an additional compression cycle, can reduce the need for backup heating in colder regions and maintain comfort on colder days.[5] Ground-source heat pumps (GSHPs), which transfer heat between indoor space and the nearby ground, can also operate more efficiently in colder climates but are less common due to their higher capital costs and specific technical skills required for installation.[6] Water source heat pumps use submerged pipes to transfer heat to and from a water source and are thus limited to areas with an appropriate source.

The costs of heat pumps vary substantially across households and regions within the U.S.[7] For air source heat pumps, typical installation costs range from $5,000 to $10,000, with an average of $7,500 (adjusted to 2024 prices).[8] Policy incentives, including through the Inflation Reduction Act (IRA), can reduce these costs. Home size, efficiency, and climate can also lead to different sizing specifications and equipment costs. Costs of labor and technician experience with recommending and installing heat pumps can also vary across locations. As demand for heat pumps increases with policy support and market incentives (and high natural gas prices driven by geopolitical events and other factors), the costs of heat pumps may also rise temporarily as value chains adjust.[9] Large regional differences in operating costs of heat pumps are also common and depend on local climate and the costs of electricity compared to natural gas and other fuels.[10] The cost landscape for heat pumps is also changing. For instance, previous studies have found that a smaller portion (~30%) of households would see net benefits when switching to a single speed heat pump[11,12], whereas more recent findings that include cold climate heat pumps find they can be cost effective for a large portion (~60%) of households.[13] This large variation in costs is a key motivation for the framework we propose that uses net benefits in assessing inequities in heat pump use across the U.S.

Several factors are likely to change the net private and social benefits of heat pumps in the future. Technological improvements will allow heat pumps to operate more efficiently (requiring less electricity to provide the same heating output), particularly in colder climates where efficiency poses the greatest barriers. New business models that lower installation costs and provide greater operational flexibility could also increase the value of heat pumps for users and electric utilities.[14] Furthermore, net benefits could change if the price of natural gas and other fuels rise or fall relative to the price of electricity. From a climate perspective, studies suggest that heat pump use rates of up to 90% can cost-effectively reduce greenhouse gas emissions in the U.S. by 2050.[12] However, potential cost setbacks could also occur, including reductions in fossil fuel prices with reduced demand and as well as challenges for electric grid operations with large-scale use.[15] Nevertheless, taken together, research suggests that future decarbonization of electricity and heat pump cost declines due to technological improvements can further increase the public and private benefits of heat pump use across the U.S.

## Section 2: Heat pump policies and incentives

There is a strong rationale for public policies to support heat pump adoption. The net public benefits (including health and environmental benefits) of heat pumps are positive in many parts of the U.S., but the private costs to households may be greater than fossil fuel alternatives (particularly natural gas).[16] Even in locations where heat pumps have lower life cycle costs, high up-front costs can impede adoption,[17] a challenge that is further exacerbated by credit and liquidity constraints for some households.[18] These costs may include the heat pump itself as well as electrical panel and other home upgrades that may be required.[19] Cost barriers may be especially acute in emergency installation cases, where the current heating system has failed and there may be limited time for homeowners to upgrade home systems to be heat pump ready or research and learn about new heating alternatives. Furthermore, in some areas, the operating costs of heat pumps exceed those of natural gas furnaces. This feature makes heat pumps different from many other household technologies such as energy efficiency or rooftop solar, where operating costs are relatively low or zero.

Policies to incentivize energy efficiency and clean energy technologies typically aim to improve access to capital, reduce the burden of high upfront costs, lower financing costs, and address split incentives.[20] At the federal level, schemes like the Weatherization Assistance Program (WAP) install energy efficiency upgrades for low-income residents, such as insulation, furnace repair or replacement, duct sealing, and refrigerator replacements.[21] State and local governments may also provide tax incentives, direct rebates, or grants. Additionally, in most states, utility regulatory commissions mandate utilities to use ratepayer dollars to fund low-income energy efficiency programs,[22] and utilities may also provide incentives through mechanisms such as on-bill financing. These approaches have also been used by various states to incentivize rooftop solar, including on-bill financing for solar installations and no-cost installation for low-income households.[23] These policies and programs continue to serve an important role in motivating residential energy efficiency and clean energy use. However, new policy mechanisms will also be important; for example, policies targeting increased adoption by landlords will be important to increasing technology use by renters.

Heat pump incentives often take the form of tax breaks, rebates, or low-/no-interest loans.[24] Many of these are administered by utilities at the behest of state regulators. For example, Duke Energy in North Carolina offers customers a rebate of several hundred dollars for replacing baseboard (i.e., electric resistance) heating with a heat pump.[25] Others are overseen by statewide organizations that facilitate rebates and incentives. For instance, Focus on Energy in Wisconsin administers rebates for customers of participating utilities.[26] Other states provide direct support for heat pumps, including northeastern states such as Maine, Vermont, and Massachusetts. Policies differ by state, but most take the form of rebates for either customers (downstream) or installers (upstream or midstream). The rebate amount is generally determined by household annual income, heat pump size, fuel type, and/or duct status. The Inflation Reduction Act (IRA) substantially expands incentive programs nationally to provide up to $14,000 in direct consumer rebates for heat pumps and other energy-efficient appliances.[27]

Most incentives are contingent upon the installation of heat pumps that meet certain performance and efficiency standards and the replacement of select fuels and equipment such as lower-efficiency electric heat or propane and fuel oil systems. Incentives can also be targeted to different home types. For instance, Energy Trust of Oregon provides rebates for single-family homeowners, but most incentives require that the heat pump replaces an electric furnace or resistance heater.[28] Additionally, to promote workforce development and control installation quality, many incentive programs, such as the New York State Clean Heat program by the New York State Energy Research and Development Authority (NYSERDA), requires that heat pumps be installed by a "trade ally" or approved contractor.[29] Other programs, such as Southern California Edison, provide incentives directly to installers to pass along to customers, effectively incentivizing customers to work with approved contractors.[30]

An growing number of incentive programs also offer additional funding and opportunities for financing for low-income and qualified households. For example, the Puget Sound Energy Efficiency Boost program previously provided up to an additional $4,000 for income-qualified homes to install a heat pump on top of their standard rebates.[31] Likewise, Vermont offers additional rebates for people below a certain income, while New Hampshire offers whole-house incentives to offset greater portions of energy use.[32] Such policies seek not only to incentivize the adoption of energy-efficient technologies like heat pumps but also to bring down household energy bills and invest in energy justice. The IRA also provides additional funding for households below 80% of the area median income (AMI).

## Section 3: Description of variables

This section describes the equity and control variables included in our analysis, including their theoretical basis, data sources, and interpretation (see Table 3.1 for descriptive statistics). The ZTRAX dataset includes roughly 164 million buildings of which 65 million are residential buildings that report a heating type. While this reduced sample raises potential concerns around missing data biases, robustness checks suggest there to be little to no bias introduced as a result of missing records (see Section 3.11 for the results of this robustness check).

### 3.1 Median Income

Income can predict household use of energy-efficient and clean energy technologies, many of which have high up-front costs. Following previous work, we use the logged median annual household income at the census tract level using table S1901 of the five-year estimates from the American Community Survey (ACS) for 2016-2020.[33–36] To further account for tail effects at the far upper and lower ends of income, we also include the square of logged income.

### 3.2 Tenancy Status

Homes are less likely to have energy efficiency and clean energy technologies if they are renter and not owner occupied, due to the split incentive problem between landlords and tenants.[34] We extract data from ACS table B25002, which classifies every housing unit as either vacant or occupied, and ACS table B25008, which divides the occupied housing units into renter-occupied and owner-occupied categories.[35] Combining both the tables, we establish the percentage of homes that are vacant, owner-occupied, or renter-occupied in each census tract. These variables are used to estimate the impact of homeownership on heat pump use.

### 3.3 Race/ethnicity

Race and ethnicity can predict use of energy efficiency and clean energy technologies, even after controlling for income and tenancy status, indicating possible systematic racial/ethnic barriers to use of these technologies. We use categories for race/ethnicity defined in ACS table B03002, which include: White, Black, Hispanic, Asian, Pacific Islander, Native, "Two-or-more," and "Other".[35] The Hispanic category refers to those identifying as only Hispanic or Latino or Hispanic or Latino along with another race category. These values are divided by the total population in each census tract to calculate the percent of the population that identifies as different racial/ethnic groups. We use white as the baseline category.

### 3.4 Building Age

Newer buildings are more likely to use electricity for space heating, whereas older buildings are more likely to heat with natural gas, oil, propane, or wood.[37] We include both the census-tract-level building age from the five-year estimates from the ACS[35] and the individual year of construction or last code-triggering renovation for each unit included in the ZTRAX database. The primary analysis utilizes the latter to develop a continuous building age variable. We also use building age for ACS as a robustness check. ACS reports building age in decade groupings and coded as dummy variables: pre-1940 construction, 1940-49, 1950-59, 1960-69, 1970-79, 1980-89, 1990-99, 2000-10, and post-2010.[38] We use new construction (post-2010) as the baseline because installing heat pumps in new buildings is widely seen as easier and less expensive compared to retrofits in older buildings.[39]

### 3.5 Electricity prices

Higher electricity prices increase the costs of operating a heat pump as well as an electric resistance heater, with costs increasing more for electric resistance heaters due to their lower operating efficiencies (compared to a heat pump). We use electricity price data from EIA form 861, which reports average prices per kWh for residential customers at the utility-county level, for the years 2016 through 2021.[40] Total revenue from residential sales by utility-county were divided by total kWh sales by utility-county to assess average price for each year. Three Texas utilities also did not report prices during the period of the analysis (Oncor, CenterPoint, and Texas-New Mexico). For homes serviced by these utilities, we use average county data from their FERC form 1 reports (page 300).[41] We also cleaned the data by reassigning county names or boundaries that had changed since the time of the last report (namely in Alaska). Prices were then normalized to 2020 prices using the EIA national electricity price index. To identify county-level prices for counties serviced by more than one utility, we calculated a customer-weighted average electricity price. Electricity prices were subsequently standardized by mean centering and dividing by their standard deviation.

Prior literature has leveraged state-level average electricity prices to assess the relationship between electricity costs and heat pump use.[42] However, due to the collinearity of other state effects (including policies) and state-level prices, such analyses may be biased. By leveraging sub-state (i.e., county-level) electricity prices, our analysis identifies the role of within-state electricity price variation in heat pump use. We note that these effects can depend strongly on the dominant heating technology because an increase in electricity price decreases the relative benefits of replacing a furnace with a heat pump but increases the relative benefits of replacing an electric resistance heater with a heat pump. We perform several robustness checks on the role of electricity prices including: (1) replicating our main analysis without electricity and gas prices (see Section 8.16), (2) performing an analysis using state-level average electricity prices for both the ZTRAX and RECs data without including state fixed effects (see Sections 6 and 8.14), and (3) replicating our main analysis separately for regions where either electricity or other fuels (including natural gas) are the dominant heating technology (see Section 8.15). These robustness checks provide additional support for analysis.

### 3.6 Natural gas prices and access

Low natural gas prices make switching to a heat pump comparatively less advantageous. Prior research shows that operational cost is a significant predictor of household decisions to install a heat pump, with natural gas price (and access, as 17% of homes do not have gas access) predicting lower heat pump use.[12] First, we map each household in ZTRAX (latitude and longitude) onto the publicly available natural gas service territories from Homeland Infrastructure Foundation-Level Data (HIFLD).[43] Second, we obtain natural gas prices for each utility boundary from the EIA-176 forms for 2016 (in USD per thousand cubic foot).[44]

Approximately 16% of territories were not matched to a record in the EIA-176 form. We used fuzzy string matching to match additional utility territories to pricing data. We matched approximately 7% of utility boundaries with strict criteria, which left approximately 9% of utilities unmatched. For the remaining utilities, we used state average prices reported from EIA[44] for 2020. We use standardized prices in our logistic regression models. In addition, we included a binomial utility gas access variable which was defined as 1 if the household was served by a natural gas utility service company and 0 otherwise and used this variable in place of gas prices as a robustness check on our model (see Section 8.8).

### 3.7 Climate

Heat pumps are less efficient in cold climates, and thus less likely to be cost effective.[45] Additionally, they may be less likely to be present in areas with minimal heating needs, where the benefits of their increased efficiency relative to electric resistance heating are less impactful compared to the higher upfront cost. We include heating and cooling degree days (HDD and CDD) in our model to control for heating and cooling requirements. For cooling degree days, the value reflects the number of days the average outside air temperature of a given location was above 65°F (the standard temperature used in the U.S. to maintain indoor comfort) multiplied by the difference between 65°F and the average daily temperature. For heating degree days, the value reflects the inverse (i.e., days with average temperature below 65°F, when heating is needed to maintain comfort). Annual HDD and CDD values with a standard reference of 65°F were obtained from the National Oceanic and Atmospheric (NOAA) 1991-2020 weather normals for 7300 weather stations nationwide.[46] We then matched each household to the nearest weather station and assigned the recorded HDD and CDD values.

## 3.8 State fixed effects

We expect some local fixed effects on heat pump use given differences in the policies, histories, geographies, markets, and trends of individual states. To control for these effects, we include state-level fixed effects in our parametric analysis. Owing to reporting differences between jurisdictions, several states also reported zero heat pumps within their service territory and were excluded from the analysis. These states include Hawaii, Louisiana, New Jersey, New York, and Vermont. Alaska was used as the reference state.

## 3.9 Housing type

To control for differences in household type, we use dummy variables for housing type designations in the ZTRAX database. We use five different housing types in our analysis: (1) Single-family, (2) Single-family shared wall, (3) Multi-family, (4) Trailer/Mobile home, and (5) Other. Multi-family was used as the reference category.

## 3.10 Descriptive statistics for variables

***Table 3.1.*** *Descriptive statistics for variables used in our parametric analysis to assess spatial patterns in heat pump use across the U.S.*

| Variable | Mean | Standard Deviation | Unit |
|---|---|---|---|
| Home Value | 330,041 | 2,153,357 | $ house price |
| Electricity price | 0.13 | 0.04 | $/kWh |
| Natural gas price | 11.94 | 3.62 | $/thousand ft3 |
| Gas Access (%) | 0.84 | 0.35 | % with access |
| Heating degree-days | 3,951 | 2,093 | °F-degree days/year |
| Cooling degree-days | 1,553 | 1,022 | °F-degree days/year |
| Median Income | 70,123 | 34,813 | $ annual income |
| Building age | 46.54 | 22.21 | years |
| Owned (%) | 62.54 | 22.37 | % homes |
| Rented (%) | 32.18 | 21.56 | % homes |
| Vacant (%) | 5.28 | 6.78 | % homes |
| White (%) | 62.25 | 28.99 | % individuals |
| Black (%) | 12.79 | 20.33 | % individuals |
| Asian (%) | 4.67 | 8.51 | % individuals |
| Hispanic (%) | 14.36 | 19.33 | % individuals |
| Multiracial (%) | 4.91 | 4.61 | % individuals |
| Pacific Islanders (%) | 0.13 | 0.75 | % individuals |
| Native (%) | 0.60 | 3.11 | % individuals |
| Other (%) | 0.29 | 1.10 | % individuals |
| Multifamily Building Type (%) | 4.26 | 10.40 | % buildings |
| Other Building Type (%) | 1.14 | 6.01 | % buildings |
| Single Family Building Type (%) | 82.27 | 23.50 | % buildings |
| Single Family Shared Wall Building Type (%) | 12.31 | 20.51 | % buildings |
| Trailer Home Building Type (%) | 0.02 | 0.61 | % buildings |

### 3.11 Missing data robustness check

The ZTRAX data includes the universe of buildings in the U.S. including both residential and commercial spaces. Of the roughly 164 million units in the dataset, only 65 million provide data on heating systems. As such, there may be concerns that the results are subject to missing data bias. Returning to the original, uncleaned data (prior to dropping entries with missing data), we calculated a Pearson's correlation coefficient matrix for the rate of missingness in each of the variables of interest (see Table 3.2). Missingness is approximated by the number of null values in the original dataset. The correlation coefficients thus represent the relationship between a missingness and each of the other variables. None of the correlation coefficients for any of the demographic variables of interest (or electricity prices) are significant, suggesting that our data cleaning process and the omission of data from our analysis do not ultimately bias our results. This alongside the analysis of the nationally-representative RECS data provides additional evidence that the findings in this analysis are not biased by missing data.

***Table 3.2.*** *Pearson correlation coefficients for relevant variables with missing heating type indicator in the ZTRAX dataset.*

| Variable | Missing heating type |
|---|---|
| White (%) | -0.053 |
| Black (%) | 0.018 |
| Asian (%) | 0.022 |
| Hispanic (%) | 0.049 |
| Pacific Islander (%) | 0.018 |
| Native (%) | 0.016 |
| Multirace (%) | 0.006 |
| Other (%) | -0.005 |
| Median income | -0.061 |
| Owned (%) | -0.096 |
| Rented (%) | 0.055 |
| Vacant (%) | 0.103 |
| Electricity price | 0.108 |

## Section 4: National regression results

***Table 4.1***. *Coefficient estimates and standard errors for the national logistic regression model. State (extracted from ZTRAX) dummy variables are also included in the model but not shown here for brevity. The percentage change in odds for a unit change in log income was calculated with a value of $70,123 (average of census-tract median income values in the dataset). The percentage change in odds for building age, heating degree-days, cooling degree-days, electricity price, and natural gas price were calculated using the corresponding mean values (μ) and standard deviations (σ) (see Section 3.10). The marking in the coefficient estimates denotes the significance of the coefficient (i.e., *** p<0.01, ** p<0.05, * p<0.1). Variables labeled "-" are dimensionless.*

| Variable | Log Odds Coefficient | S.E. | Unit Change | Odds Change (%) |
|---|---|---|---|---|
| Black (%) | -1.442*** | 0.005 | 0.1 | -13.429 |
| Asian (%) | -4.308*** | 0.017 | 0.1 | -35.001 |
| Hispanic (%) | -0.906*** | 0.01 | 0.1 | -8.662 |
| Multiracial (%) | -2.801*** | 0.025 | 0.1 | -24.429 |
| Pacific Islanders (%) | -8.156*** | 0.174 | 0.1 | -55.763 |
| Native (%) | 1.295*** | 0.047 | 0.1 | 13.826 |
| Other (%) | -5.779*** | 0.098 | 0.1 | -43.892 |
| Log Income (-) | 3.595*** | 0.072 | Log(1.1) | -4.711 |
| (Log Income)$^2$ (-) | -0.183*** | 0.003 | - | - |
| Rented (%) | -0.428*** | 0.007 | 0.1 | -4.19 |
| Vacant (%) | 0.961*** | 0.011 | 0.1 | 10.087 |
| Log Home value (-) | 0.016*** | 0.001 | Log(1.1) | 0.155 |
| Building age (-) | -0.741*** | 0.001 | 0.1μ/σ | -14.382 |
| Heating degree-days (-) | -0.523*** | 0.005 | 0.1μ/σ | -9.401 |
| Cooling degree-days (-) | -0.172*** | 0.005 | 0.1μ/σ | -2.58 |
| Electricity price (-) | 0.217*** | 0.002 | 0.1μ/σ | 7.307 |
| Natural gas price (-) | -0.022*** | 0.001 | 0.1μ/σ | -0.720 |
| Multifamily (0,1) | -0.246*** | 0.007 | 1 | -21.808 |
| Shared-wall single-family (0,1) | 0.050*** | 0.002 | 1 | 5.085 |
| Mobile home (0,1) | 0.183*** | 0.059 | 1 | 20.081 |
| Other housing (0,1) | 0.048*** | 0.011 | 1 | 4.917 |
| State dummies | Yes | - | - | - |
| Constant | -22.69*** | 0.407 | - | - |
| Pseudo $R^2$ | 0.3333 | - | - | - |
| Observations | 51,139,213 | - | - | - |

## Section 5: Regional regression results

***Table 5.1***. *Coefficient estimates and standard errors for the regional logistic regression models for heat pump use, where regions are defined based on household benefits (in the form of reduced energy bills). State (extracted from ZTRAX) dummy variables are also included in the model but not shown here for brevity. The percentage change in odds for a unit change in log income was calculated with a value of $70,123 (average of census-tract median income values in the dataset, see Figure 5 in the main paper for values in percentage change in odds). The marking in the coefficient estimates denotes the significance of the coefficient (i.e., *** p<0.01, ** p<0.05, * p<0.1). Variables labeled "-" are dimensionless.*

| Variable | Net benefits | | Neutral | | Net costs | |
|---|---|---|---|---|---|---|
| | Log Odds | S.E. | Log Odds | S.E. | Log Odds | S.E. |
| Black (%) | -1.401*** | 0.005 | -2.982*** | 0.037 | -0.523*** | 0.026 |
| Asian (%) | -5.519*** | 0.022 | -2.851*** | 0.064 | -0.082*** | 0.028 |
| Hispanic (%) | -0.879*** | 0.011 | -1.119*** | 0.035 | -0.288*** | 0.029 |
| Multiracial (%) | -2.819*** | 0.027 | -2.948*** | 0.090 | -0.951*** | 0.075 |
| Pacific Islanders (%) | -9.355*** | 0.202 | -0.298 | 0.355 | 4.920*** | 0.846 |
| Native (%) | 0.029 | 0.061 | 1.913*** | 0.138 | -0.128 | 0.217 |
| Other (%) | -6.462*** | 0.112 | -2.857*** | 0.326 | -0.321 | 0.300 |
| Log Income (-) | 0.396*** | 0.080 | 2.471*** | 0.269 | 10.52*** | 0.256 |
| $(\text{Log Income})^2$ (-) | -0.032*** | 0.004 | -0.122*** | 0.012 | -0.492*** | 0.011 |
| Rented (%) | -0.131*** | 0.007 | 0.307*** | 0.023 | -2.133*** | 0.022 |
| Vacant (%) | 1.192*** | 0.012 | 2.085*** | 0.043 | -0.511*** | 0.090 |
| Log Home value (-) | -0.026*** | 0.001 | 0.161*** | 0.002 | 0.076*** | 0.002 |
| Building age (-) | -0.822*** | 0.001 | -0.601*** | 0.004 | -0.162*** | 0.003 |
| Heating degree-days (-) | -0.215*** | 0.006 | 0.188*** | 0.019 | -1.569*** | 0.030 |
| Cooling degree-days (-) | 0.066*** | 0.005 | 0.751*** | 0.018 | -1.795*** | 0.028 |
| Electricity price (-) | 0.360*** | 0.003 | -0.012* | 0.008 | -0.937*** | 0.012 |
| Natural gas price (-) | -0.079*** | 0.002 | 0.010 | 0.008 | -0.168*** | 0.009 |
| Multifamily (0,1) | -0.284*** | 0.009 | 0.211*** | 0.013 | -0.584*** | 0.030 |
| Shared-wall single-family (0,1) | -0.028*** | 0.002 | 0.683*** | 0.007 | 0.449*** | 0.006 |
| Mobile home (0,1) | 0.055 | 0.058 | omitted | – | -2.344** | 1.004 |
| Other housing (0,1) | 0.132*** | 0.013 | -1.205*** | 0.056 | 0.926*** | 0.037 |
| State dummies | Yes | - | Yes | - | Yes | - |
| Constant | -6.626*** | 0.448 | -22.15*** | 1.514 | -61.29*** | 1.493 |
| Pseudo $R^2$ | 0.3189 | - | 0.1953 | - | 0.2642 | - |
| Observations | 27,216,988 | - | 14,454,856 | - | 9,466,140 | - |

## 5.1 Summary statistic across benefits regions

Table 5.2 shows the mean values for relevant variables across each of the benefit zones. Notably, net benefit regions tend to be warmer (higher CDD relative to HDD), have a higher ratio of natural gas to electricity prices, and have higher Black and Hispanic populations. In Table 5.3, we then show the percentage of each race/ethnicity across benefits zones. Here, we observe that the majority of the Black population lives in the net benefits region, while the majority of the Asian, Hispanic, Pacific Islander, Multi-race, and other demographic groups live in the neutral zone. The plurality of White and Native Americans live in the neutral region.

***Table 5.2**. Mean values for relevant variables (HDD/CDD and energy prices) and estimated population composition of each benefit region. Population estimates were obtained from American Community Survey 5-Year Data (2015-2020).*

| Variable | Net benefits | Neutral | Net costs |
|---|---|---|---|
| HDD (°F-degree days/year) | 3,016 | 4,942 | 6,132 |
| CDD (°F-degree days/year) | 1,969 | 1,040 | 933 |
| Electricity price ($/kWh) | 0.12 | 0.14 | 0.16 |
| Natural gas price ($/thousand sq. ft.) | 12.76 | 11.84 | 9.01 |
| White (%) | 59.05 | 58.29 | 76.31 |
| Black (%) | 15.88 | 10.14 | 8.82 |
| Asian (%) | 3.39 | 7.42 | 3.34 |
| Hispanic (%) | 18.19 | 20.21 | 7.59 |
| Pacific Islander (%) | 0.10 | 0.16 | 0.05 |
| Native (%) | 0.58 | 0.58 | 0.88 |
| Multirace (%) | 2.53 | 2.85 | 2.76 |
| Other (%) | 0.28 | 0.35 | 0.24 |

***Table 5.3**. Percentage of each racial group residing in each benefit region. Population estimates were obtained from American Community Survey 5-Year Data (2015-2020).*

| Variable | Net benefits | Neutral | Net costs |
|---|---|---|---|
| White (%) | 39.18 | 49.09 | 11.73 |
| Black (%) | 51.55 | 41.81 | 6.64 |
| Asian (%) | 24.97 | 69.33 | 5.70 |
| Hispanic (%) | 39.89 | 56.25 | 3.86 |
| Pacific Islander (%) | 32.62 | 63.66 | 3.72 |
| Native (%) | 38.11 | 48.43 | 13.46 |
| Multirace (%) | 37.26 | 53.31 | 9.43 |
| Other (%) | 35.39 | 57.40 | 7.21 |
| Total population (%) | 39.98 | 50.75 | 9.26 |

### 5.2 Significance correction

Table 5.4 shows the corrected p-values using the Holm-Bonferroni method for multiple hypothesis testing. We use a family-wise error rate (α) equal to 0.05. For the national case, all predictors are significant, even after correcting for significance for multiple hypotheses given the large number of predictors in our analysis. For the region of the net benefits, mobile home and Native are not significant. For the neutral region, electricity and natural gas prices, as well as Pacific Islanders are not significant. For the net costs region, mobile home, Native and other race are not significant. We omit variables with corrected p-values equal to zero (i.e., significant) for brevity, considering three significant figures.

***Table 5.4***. *Non-zero corrected p-values using the Holm-Bonferroni method. Family-wise error rate is equal to 0.05. Corrected p-values equal to zero are omitted.*

| Variable | National | Net benefits | Neutral | Net costs |
|---|---|---|---|---|
| Mobile home | 0.004 | 1 | 0 | 0.12 |
| Native | 0 | 1 | 0 | 1 |
| Electricity price | 0 | 0 | 0.294 | 0 |
| Natural gas price | 0 | 0 | 0.422 | 0 |
| Pacific Islanders | 0 | 0 | 0.422 | 0 |
| Other race | 0 | 0 | 0 | 1 |

## Section 6: Replication results

As with many previous studies investigating equity in clean energy technologies,[34,47] our study evaluates inequities at the census tract level. However, it is possible that this approach can under- or overestimate inequities (for example, if Black households in a location with a higher proportion of white residents are more likely to use heat pumps than Black households in a location with a higher proportion of Black residents). To explore this possibility, we replicate our analysis using the 2020 Residential Energy Consumption Survey (RECS).[48] RECS is a survey conducted by the U.S. Energy Information Administration (EIA) that includes household-level variables such as geography, climate, income, household quality, race/ethnicity, and renter status. The 2020 RECS sample was designed to meet precision requirements on energy consumption for all 50 states and Washington, DC, with a total of 18,496 respondents.[49] We use RECS data to assess how the use of heat pumps as the main heating equipment correlates with equity (race/ethnicity, income, and renter status) and other control variables at the household level. The model is implemented in R using the survey package,[50] recommended by EIA to capture the replicate weights used in the RECS methodology.[51]

The results for the relationship between household race/ethnicity and heat pump use are generally consistent with the results from our main analysis (see Tables 6.1 and 6.2). Specifically, heat pump use is significantly lower among Black or African American households than white ones, and among households of Hispanic and Latino descent than non-Hispanic/Latino households (RECS codes for Hispanic/Latino as a separate ethnicity variable). This result remains robust when controlling for additional factors: income, renter status, building type and year, climate, natural gas access, and state (Table 6.1). The direction of the relationship for other racial categories (except American Indian or Alaska Native) is also negative, consistent with our main analysis, although it is not statistically significant. In a model where we included natural gas and electricity prices (Table 6.2), all racial categories are negatively correlated with heat pump use. Overall, these results suggest that the disparities observed in our analysis are robust and likely not due to systematic biases in ZTRAX.

The following list elaborates on the variables included in the RECS analysis, their interpretability, and key differences from the ZTRAX dataset variables, if any:

1. ***Race:*** The racial categories in the RECS are (1) White alone, (2) Black or African American alone, (3) American Indian or Alaska Native alone, (4) Asian alone, (5) Native Hawaiian or other Pacific Islander alone, and (6) multiracial. As in the main analysis, white is used as the baseline group.
2. ***Ethnic Descent:*** RECS delineates Hispanic/Latino descent as a separate variable from race. This binary variable is added to all logit models, with 0 referring to non-Hispanic/Latino and 1 referring to Hispanic/Latino.
3. ***Income:*** While the median annual income (at the census tract level) in the primary analysis is a continuous variable, RECS groups income into 16 ranges, from less than $5,000 to $150,000 or more. We use the smallest range as the baseline group.
4. ***Building Age:*** In RECS, building age is grouped into five or ten year ranges from 1950 to 2020, with an additional pre-1950 category, whereas this variable is continuous in our main analysis. The age range closest to present is used as the baseline group.
5. ***Housing Type:*** There are five building types in RECS: (1) mobile home, (2) single-family house detached from any other house, (3) single-family house attached to one or more other houses (for example, duplex, row house, or townhome), (4) apartment in a building with 2 to 4 units, and (5) apartment in a building with 5 or more units. As in the main analysis, detached single-family house is used as the baseline group.
6. ***Natural Gas Price and Access:*** RECS includes a dummy variable for natural gas access. We use 1 to refer to a household that has natural gas available in underground pipes, and 0 otherwise. In another model, we include state level natural gas prices from EIA and remove the natural gas access variable.[52]
7. ***State Level Electricity Price:*** Electricity prices from the EIA reported at the state level and included in one of our models. We excluded state fixed effects in models with state electricity prices to avoid collinearity issues.

8. ***Other Variables:*** The remaining variables in RECS are largely similar to our main analysis, with minor differences. Tenancy in RECS includes an additional status of occupancy without payment of rent.

***Table 6.1.*** *Coefficient estimates and standard errors of the logistic regression model using the RECS dataset (and including natural gas access). Symbols next to coefficient estimates denote significance (i.e., *** p<0.01, ** p<0.05, * p<0.1). (HH = household.) Variables labeled "-" are dimensionless.*

| Variable | Log Odds Coefficient | S.E. | Unit Change | Odds (%) |
|---|---|---|---|---|
| Black or African American HHs | -0.259*** | 0.093 | 0.1 | -2.556 |
| American Indian or Alaska Native HHs | 0.088 | 0.301 | 0.1 | 0.881 |
| Asian HHs | -0.108 | 0.147 | 0.1 | -1.077 |
| Native Hawaiian or Other Pacific Islander HHs | -0.211 | 0.615 | 0.1 | -2.093 |
| 2 or more Race HHs | -0.147 | 0.201 | 0.1 | -1.461 |
| Hispanic or Latino Descent HHs | -0.469*** | 0.106 | 0.1 | -4.583 |
| Annual HH Income $5,000 - $7,499 | -0.172 | 0.288 | 1 | -15.835 |
| Annual HH Income $7,500 - $9,999 | 0.155 | 0.266 | 1 | 16.822 |
| Annual HH Income $10,000 - $12,499 | -0.204 | 0.237 | 1 | -18.430 |
| Annual HH Income $12,500 - $14,999 | -0.330 | 0.263 | 1 | -28.075 |
| Annual HH Income $15,000 - $19,999 | 0.200 | 0.207 | 1 | 22.189 |
| Annual HH Income $20,000 - $24,999 | 0.079 | 0.193 | 1 | 8.207 |
| Annual HH Income $25,000 - $29,999 | -0.053 | 0.201 | 1 | -5.131 |
| Annual HH Income $30,000 - $34,999 | 0.011 | 0.188 | 1 | 1.143 |
| Annual HH Income $35,000 - $39,999 | -0.046 | 0.200 | 1 | -4.526 |
| Annual HH Income $40,000 - $49,999 | 0.009 | 0.180 | 1 | 0.921 |
| Annual HH Income $50,000 - $59,999 | 0.072 | 0.179 | 1 | 7.441 |
| Annual HH Income $60,000 - $74,999 | 0.148 | 0.174 | 1 | 15.934 |
| Annual HH Income $75,000 - $99,999 | 0.128 | 0.172 | 1 | 13.628 |
| Annual HH Income $100,000 - $149,999 | 0.125 | 0.174 | 1 | 13.346 |
| Annual HH Income $150,000 or more | 0.158 | 0.175 | 1 | 17.155 |
| Rented HHs | 0.028 | 0.084 | 0.1 | 0.279 |
| Occupied Without Payment of Rent | -0.253 | 0.248 | 0.1 | -2.495 |
| Year Built 2010 to 2015 | -0.175 | 0.158 | 0.1 | -1.738 |
| Year Built 2000 to 2009 | -0.345** | 0.134 | 0.1 | -3.388 |
| Year Built 1990 to 1999 | -0.244 | 0.136 | 0.1 | -2.407 |
| Year Built 1980 to 1989 | -0.226 | 0.139 | 0.1 | -2.237 |
| Year Built 1970 to 1979 | -0.347** | 0.140 | 0.1 | -3.414 |
| Year Built 1960 to 1969 | -0.436*** | 0.151 | 0.1 | -4.265 |
| Year Built 1950 to 1959 | -0.549*** | 0.160 | 0.1 | -5.342 |
| Year Built Before 1950 | -0.745*** | 0.156 | 0.1 | -7.175 |
| Heating degree-days (-) | 0.000*** | 0.000 | 1 | -0.041 |
| Cooling degree-days (-) | 0.000*** | 0.000 | 1 | -0.015 |
| Natural Gas Access | -1.161*** | 0.058 | 1 | -68.686 |
| Housing Type, Single Shared Wall (0,1) | -0.227** | 0.104 | 1 | -20.308 |
| Housing Type, Multi 2 to 4 Units (0,1) | 0.107 | 0.139 | 1 | 11.293 |
| Housing Type, Multi 5 or more Units (0,1) | -0.078 | 0.109 | 1 | -7.504 |
| Housing Type, Mobile (0,1) | -0.084 | 0.112 | 1 | -8.057 |
| State dummies | Yes | - | - | - |
| Constant | 1.562*** | 0.305 | - | - |
| Pseudo $R^2$ | 0.211 | - | - | - |
| Observations | 18,496 | - | - | - |

***Table 6.2.*** *Coefficient estimates and standard errors of the logistic regression model using the RECS dataset (and including state-level electricity and natural gas prices). The marking in the coefficient estimates denotes the significance of the coefficient (i.e., *** p<0.01, ** p<0.05, * p<0.1). (HH = household.) Variables labeled "-" are dimensionless.*

| Variable | Log Odds Coefficient | S.E. | Unit Change | Odds (%) |
|---|---|---|---|---|
| Black or African American HHs | -0.222** | 0.087 | 0.1 | -2.196 |
| American Indian or Alaska Native HHs | -0.103 | 0.284 | 0.1 | -1.025 |
| Asian HHs | -0.304** | 0.140 | 0.1 | -2.994 |
| Native Hawaiian or Other Pacific Islander HHs | -0.641 | 0.628 | 0.1 | -6.209 |
| 2 or more Race HHs | -0.162 | 0.185 | 0.1 | -1.607 |
| Hispanic or Latino Descent HHs | -0.571*** | 0.100 | 0.1 | -5.550 |
| Annual HH Income $5,000 - $7,499 | -0.194 | 0.264 | 1 | -17.634 |
| Annual HH Income $7,500 - $9,999 | 0.127 | 0.250 | 1 | 13.542 |
| Annual HH Income $10,000 - $12,499 | -0.173 | 0.223 | 1 | -15.886 |
| Annual HH Income $12,500 - $14,999 | -0.246 | 0.246 | 1 | -21.808 |
| Annual HH Income $15,000 - $19,999 | 0.175 | 0.194 | 1 | 19.125 |
| Annual HH Income $20,000 - $24,999 | 0.080 | 0.181 | 1 | 8.329 |
| Annual HH Income $25,000 - $29,999 | -0.114 | 0.191 | 1 | -10.774 |
| Annual HH Income $30,000 - $34,999 | 0.003 | 0.180 | 1 | 0.300 |
| Annual HH Income $35,000 - $39,999 | -0.117 | 0.187 | 1 | -11.041 |
| Annual HH Income $40,000 - $49,999 | 0.006 | 0.170 | 1 | 0.602 |
| Annual HH Income $50,000 - $59,999 | -0.026 | 0.168 | 1 | -2.566 |
| Annual HH Income $60,000 - $74,999 | 0.065 | 0.164 | 1 | 6.716 |
| Annual HH Income $75,000 - $99,999 | 0.032 | 0.163 | 1 | 3.252 |
| Annual HH Income $100,000 - $149,999 | -0.001 | 0.163 | 1 | -0.100 |
| Annual HH Income $150,000 or more | -0.065 | 0.165 | 1 | -6.293 |
| Rented HHs | 0.017 | 0.078 | 0.1 | 0.170 |
| Occupied Without Payment of Rent | -0.144 | 0.259 | 0.1 | -1.430 |
| Year Built 2010 to 2015 | -1.038*** | 0.145 | 0.1 | -9.859 |
| Year Built 2000 to 2009 | -0.831*** | 0.149 | 0.1 | -7.974 |
| Year Built 1990 to 1999 | -0.651*** | 0.139 | 0.1 | -6.303 |
| Year Built 1980 to 1989 | -0.497*** | 0.128 | 0.1 | -4.849 |
| Year Built 1970 to 1979 | -0.253** | 0.127 | 0.1 | -2.498 |
| Year Built 1960 to 1969 | -0.276** | 0.125 | 0.1 | -2.722 |
| Year Built 1950 to 1959 | -0.336*** | 0.123 | 0.1 | -3.304 |
| Year Built Before 1950 | -0.186 | 0.147 | 0.1 | -1.843 |
| Heating degree-days (-) | 0.001*** | 0.000 | 1 | 0.100 |
| Cooling degree-days (-) | 0.000*** | 0.000 | 1 | 0.000 |
| 2020 Electricity Prices (-) | -0.204*** | 0.010 | 1 | -18.454 |
| 2020 Natural Gas Prices (-) | 0.044*** | 0.009 | 1 | 4.498 |
| Housing Type, Single Shared Wall (0,1) | -0.205** | 0.104 | 1 | -18.535 |
| Housing Type, Multi 2 to 4 Units (0,1) | 0.061 | 0.138 | 1 | 6.290 |
| Housing Type, Multi 5 or more Units (0,1) | -0.160 | 0.106 | 1 | -14.786 |
| Housing Type, Mobile (0,1) | -0.052 | 0.108 | 1 | -5.067 |
| State dummies | Yes | - | - | - |
| Constant | -1.710*** | 0.192 | - | - |
| Pseudo $R^2$ | 0.211 | - | - | - |
| Observations | 18,275 | - | - | - |

## Section 7: Estimation of heat pump benefits

This section describes our approach to estimating net energy bill savings from switching to a heat pump relative to the current primary heating fuel for each county in the U.S.

### 7.1 Data sources

We extract heating technology data from the ACS five-year survey 2016-2020.[35] Five primary fuels are used: (1) utility gas, (2) electricity, (3) bottled, tank, or liquified propane (LP) gas, (4) fuel oil or kerosene, and (5) wood. We use Typical Meteorological Year (TMY3) weather data from NOAA[53] (reported as hourly temperatures at the county level) and estimate county-level heating degree hours using a standard reference temperature of 18.3°C (65°F), which is also used by NOAA to calculate heating and cooling degree days.[46]

We use energy prices for 2020 at the state level (for natural gas and other fuels) and at the county level (for electricity). For natural gas, we use annual state average residential prices reported by EIA (in USD per thousand cubic feet)[52] and convert them to USD/kWh using the state-average heat content of utility gas delivered to customers (in BTU per cubic foot) from EIA.[54] For other fuels, we use residential pricing data inferred by the State Energy Data System (SEDS)[55] at the state level. For fuel oil, we use the prices reported for kerosene. For bottled, tank, or LP gas, we use the prices provided for propane.

For electricity prices, EIA form 861 gathers electricity prices at the utility level across the US.[40] Electricity prices are population-weighted averages per county inferred from EIA form 861 as the ratio of utility sales (USD) and total energy sold (kWh). For our calculation of net energy bill impacts, we assume that counties that currently report electricity as the primary heating source in ACS use electric resistance heating as the main technology.

### 7.2 Heating load

We use the heating degree-day method to estimate the annual heating load of a typical single-family detached house.[17] This method uses local heating degree day data (at the hourly resolution) and a heat loss factor to calculate energy demand for heating, which is the product of the heat transfer coefficient *U* and the area *A*. We focus our analysis on the sign (positive or negative) of net energy cost savings of using a heat pump in a given location, relative to the current primary heating technology in that location. Housing characteristics such as insulation and type of construction can substantially impact heating load. However, heating load affects the costs of both heat pumps and the comparison technology, and thus differences in heating load are less likely to change the sign of the net benefits of switching to a heat pump.

We use the RECS 2015 dataset to infer an average *UA* factor for a single-family, detached house in the U.S.[56] We filter RECS data to include only single-family detached houses, which includes 3,752 houses. The RECS dataset includes the average total energy use for space heating ($Q_{avg}$, 41.69 million BTU per year) and average annual heating degree days ($HDD_{avg}$, 3,773°F-day, with a reference of 65°F). We calculate the *UA* factor as follows:

$$UA = \frac{Q_{avg}}{HDD_{avg}} = \frac{41.69\times10^{6}\,BTU}{3773\,°F\cdot day},$$

which is 0.243 kWh/hr/°C. This is 7% above the typical value for a single-family detached house in the Midwest[17] (*UA* = 0.226 kWh/hr/°C), since U.S. houses, on average, will have poorer insulation. We then estimate the annual heating load ($Q_i$) for $N_c$ counties as follows:

$$Q_i = \sum_{j=1}^{8760} UA \cdot HDH_{i,j}, \quad i = 1, 2, ..., N_c\,,$$

where $HDH_{i,j}$ are the county heating degree hours, *j* are hours in the year, and *i* are counties.

### 7.3 Energy costs

***Heat pump.*** The coefficient of performance (*COP*) for a heat pump can be described as a linear function of temperature (*T*). Following previous research,[11] we calculate the parameters in this function using a range of heat pumps with a capacity of approximately 25,000 BTU/hr from a database compiled by the Northeast Energy Efficiency Partnership (NEEP).[57] (While the database has been updated, these updates do not change the coefficients substantially.) The COP is expressed as a function of outside air temperature (*T* in °F) for all counties as follows:

$$COP(T) = 0.05 \cdot T + 1.6, \quad T \geq 5\,°F\,.$$

Data for temperatures below 5°F is not available in the NEEP database. For a single-stage heat pump, it is common for the COP to fall rapidly to 1 below 5 °F (equivalent to operating as an electric resistance heater).[15] However, a cold climate heat pump, typically with two-stage compressors, can maintain a higher COP in colder temperatures.[10] We conservatively assume that the majority of heat pump installations are single-stage and rely on resistive heating when the temperature falls below 5°F. We then calculate the heat pump energy costs $h_i$ as follows:

$$h_i = Pe_i \cdot UA \cdot \sum_{j=1}^{8760} \frac{HDH_{i,j}}{COP(T_{i,j})}, \quad i = 1, 2, \ldots, N_c,$$

where $Pe_i$ is the electricity price and $T_{i,j}$ is the typical meteorological year temperature at hour $j$ and county $i$.

***Natural gas.*** Energy costs in counties where natural gas is the primary heating fuel are calculated with state average residential prices and heating content of natural gas delivered to residential customers. We assume a natural gas furnace with average efficiency (90%, based on the National Residential Efficiency Measures Database)[8] and calculate the annual cost ($c_i$) as follows:

$$c_i = Q_i \cdot \frac{Pg_i}{HC_i \cdot \eta_g}, \quad i = 1, 2, \ldots, N_c,$$

where $Q_i$ is the annual heating load (in kWh), $Pg_i$ is the utility gas price (in USD/thousand cubic foot), $HC_i$ is the average heating content (in kWh/thousand cubic foot), and $\eta_g$ is the natural gas furnace efficiency for a county *i*.

***Electric resistance.*** Energy costs in counties where electricity is the primary fuel (using an electric resistance heater) are calculated as follows:

$$c_i = Q_i \cdot Pe_i \quad i = 1, 2, \ldots, N_c,$$

where $Q_i$ is the annual heating load (in kWh) and $Pe_i$ is the electricity price (in USD/kWh).

***Other delivered fuels.*** Energy costs in counties where other heating fuels (i.e., fuel oil, propane, or wood) are the primary heating source were calculated using reported prices by the U.S. State Energy Data System (SEDS) in USD/kWh.[58] For propane and fuel oil furnaces, we use a heating device efficiency of 88%, which is the average efficiency found in the National Residential Efficiency Measures Database. For wood space heaters, we use an average efficiency of 79% for wood central heaters reported by the EPA-certified wood stove database.[59] Energy costs ($c_i$) are calculated as follows:

$$c_i = Q_i \frac{Po_i}{\eta_o},$$

where $Q_i$ is the annual heating load (in kWh), $Po_i$ is the other fuels price (in USD/kWh per county), and $\eta_o$ is the heating device efficiency of other fuels.

### 7.4 Energy savings

We estimate the net energy bill savings from switching to a heat pump from the current primary heating fuel across the U.S. at the county scale. An economic lifetime of 25 years and a private discount rate of 7% is assumed to estimate lifetime savings or added costs in our base case (see below for a discussion of our sensitivity analysis). Therefore, the lifetime savings or added costs are calculated, as the total present value, as follows:

$$s_i = \left(c_i - h_i\right) \cdot \left(1 - (1 + r)^{-n}\right)/r, \quad i = 1, 2, \ldots, N_c,$$

where $s_i$ are the lifetime fuel cost when switching to a heat pump in a county *i*, $c_i$ is the current annual energy cost, and $h_i$ is the annual energy of a heat pump, $n$ are the economic lifetime years, and *r* is the discount rate.

### 7.5 Uncertainty analysis

Many inputs that impact the net bill savings of heat pumps relative to other heating technologies are subject to uncertainty, including energy prices and device efficiencies. To assess the effects of these uncertainties on the sign of the net benefits, we use a Monte Carlo simulation. We use the following probability distributions for parameters in the analysis (with the exception of typical meteorological data, which for simplicity is deterministic in our analysis):

- ***Heat pump efficiency.*** We use a triangular distribution for the slope of the heat pump COP function with minimum, average, and maximum slope values of 0.03, 0.05, and 0.07, respectively, taken from the NEEP database. This yields a COP no higher than 5 and no less than 3 for an operating temperature of 47°F. The COP value can fall below 3 for operating temperatures lower than 47°F.
- **Other heating device efficiencies.** For natural gas, propane, fuel oil, and wood heaters, we use triangular distributions for device efficiencies. We use a minimum efficiency of 78% for natural gas, propane and fuel oil based on the first Federal standard (established in 1992)[60] for residential furnaces and 58% for wood (based on the EPA-Certified wood stove database). We use an average efficiency of 80% for natural gas and propane furnaces, 83% for fuel oil furnaces (based on the 2015 Federal standard for residential furnaces), and 79% for wood space heaters (based on the average efficiency in the EPA-certified wood stove database). We use a maximum efficiency of 99% for natural gas and propane, 96.7% for fuel oil furnaces (based on the highest efficiency Energy Star certified furnaces database),[61] and 90% for wood space heaters (based on the maximum efficiency in the EPA-certified wood stove database).
- **Fuel prices.** We use national average residential fuel prices from 2000 to 2020 reported by EIA,[62] relative to 2020, to calculate uncertainty ranges for fuel prices at the state (natural gas, propane, fuel oil, and wood) and county (electricity) level. We calculate the coefficient of variation in the time range for each fuel and use it to construct triangular distributions by setting an uncertainty range as a percentage from 2020 prices in each county. The price coefficients of variation using two standard deviations are: 30% for natural gas, 29% for electricity, 21% for propane, 79% for fuel oil, and 68% for wood.

We simulate 10,000 random samples for our Monte Carlo simulation and separate counties into three categories: (1) *net benefits*, if more than 90% of realizations yield energy savings, (2) *neutral*, when energy savings occur less than 90% and more than 10% of realizations, and (3) *net costs*, if energy savings occur in less than 10% of realizations. The results include 1749, 736, and 623 counties in the net benefits, neutral, and net costs zones.

## 7.6 Installation costs

We focus on the net benefis of heat pump operation (in the form of changes in energy bills). However, installation costs may also affect the net benefits of new heat pump adoption. We compare the costs of installing a heat pump to conventional heating technologies. We construct random triangular distributions using data from the National Residential Efficiency Measures Database (adjusted to $2020 USD)[8], which contains minimum, typical, and maximum installation costs for the U.S. While this data is only available as a national average, it includes the cost of labor and electrical components (e.g., wiring). We compare the installation costs of a heat pump against natural gas, propane, and oil furnaces and an electric resistance (baseboard) heater (see Figure 7.1). Since a heat pump can function both as a heating and cooling device, we compare the costs of each heating device plus the cost of installing a central air conditioning (AC) system.[1] The heat pump, heating devices, and air conditioning capacities are assumed to be 36,000 BTU/hr, the typical capacity in U.S. households.[63]

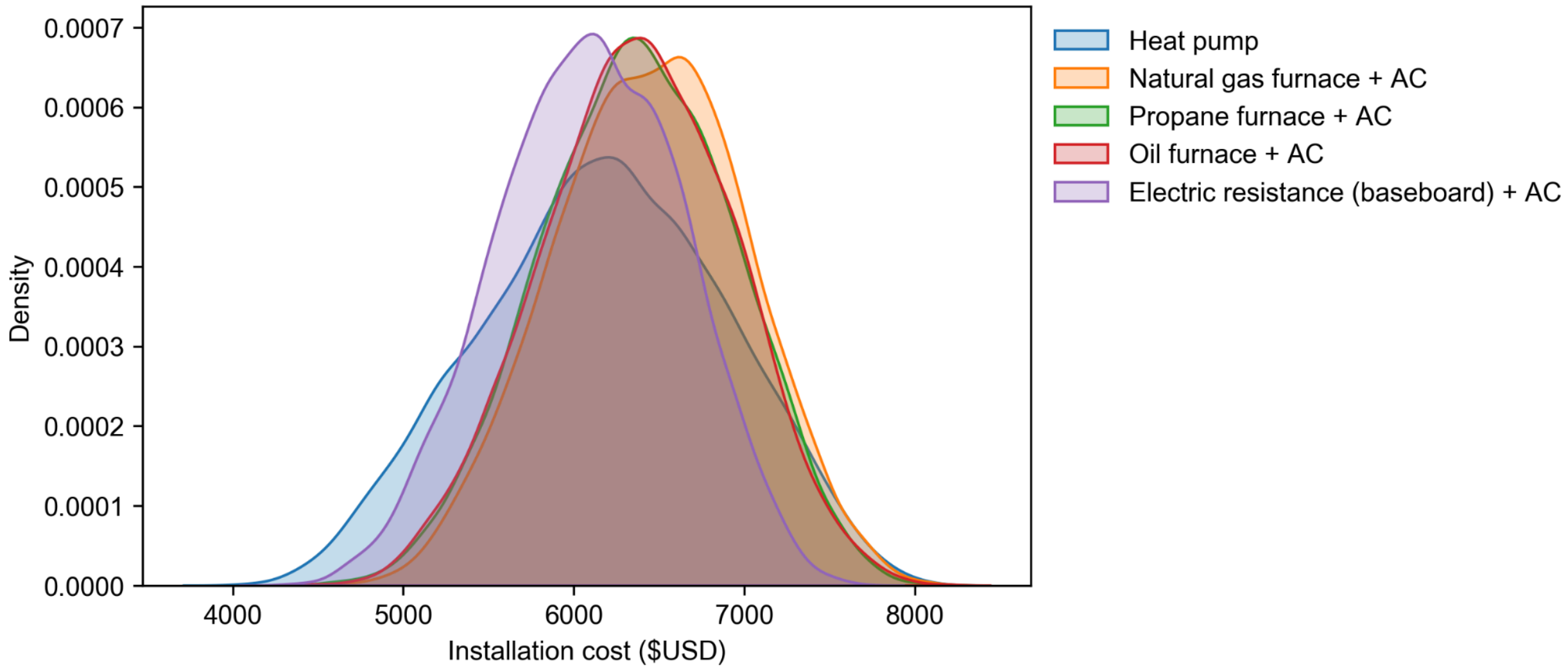

***Figure 7.1.*** *Distributions of installation costs across heating devices and central air conditioners. Data from the National Residential Efficiency Measures Database.[8]*

We observe similar distributions when comparing the costs of a heat pump to other heating systems plus the cost of a central AC system. The median installation cost of a heat pump is

[1] For standardization, other heat pump systems (e.g., mini-split) are left out of the analysis.

$6,185 with a standard deviation of $714 (see Table 7.1 for installation costs of all systems). Compared to the median of the most expensive system, a natural gas furnace plus AC, net installation savings would be only $270 on average. Compared to the least expensive system, electric resistance plus AC, net installation costs would be $90 on average. These numbers are small compared to the average operational savings of a heat pump (even given substantial regional variation in sign) of $3,300 (with an average savings of $7,144 in net benefit regions and an average cost of $2,700 in net cost regions). Furthermore, more than 80% of installation cost distributions for other sytems overlap with the cost distribution for a heat pump.

***Table 7.1.*** *Median installation costs of heating and cooling systems in Figure 7.1*

| System | Median installation cost ($USD) | Standard deviation ($USD) | Overlap % with heat pump |
|---|---|---|---|
| Heat pump | 6,185 | 714 | 0 |
| Central air conditioner (AC) | 4,684 | 519 | 0 |
| Natural gas furnace + AC | 6,455 | 559 | 82 |
| Propane furnace + AC | 6,375 | 560 | 84 |
| Oil furnace + AC | 6,375 | 560 | 84 |
| Electric resistance (baseboard) + AC | 6,095 | 540 | 84 |

## Section 8: Additional regression results

We perform several robustness checks for our analysis. We first summarize the results of these robustness checks and then provide tables with the results. Generally, these robustness checks (in addition to our replication results using the Residential Energy Consumption Survey [RECS] presented in Section 6) provide evidence that our analysis and policy takeaways are largely robust to details or potential limitations of our data and modeling approach.

We run a series of analyses where we sequentially add variables to our model. We begin with a basic national model that includes race/ethnicity, climate, energy prices, and state dummies (Section 8.1) and sequentially add income (Section 8.2), building age (Section 8.3), tenancy status (Section 8.4), home value (Section 8.5), and housing type (Section 8.6). We also run an additional analysis that replaces the continuous building age variable with a categorical variable for buildings built each decade from pre-1940s to the 2000s from the American Community Survey (Section 8.7) and an analysis that replaces natural gas price with a dummy variable for natural gas access (Section 8.8). We also run analyses that consider interaction terms for race/ethnicity and income (Section 8.9), race/ethnicity and tenancy status (Section 8.10), race/ethnicity and building age (Section 8.11), and race/ethnicity and income, tenancy status, and building age simultaneously (Section 8.12). These alternative model specifications do not change the direction of our key results on race/ethnicity, income, and renter status, except in cases where the direction of main coefficients changes due to the introduction of interaction terms (e.g., interaction with Hispanic, Multiracial, and other race in Sections 8.9 and 8.12).

We consider the inclusion of state fixed effects in our model to be important given the potential influence of state factors (including policies) on heat pump use. However, we also perform logistic regressions without state dummy variables (see Section 8.13) and with state-level rather than county-level electricity and natural gas prices (see Section 8.14). Notably, this model has a lower pseudo $R^2$ value (0.125 compared to 0.333 for our main model). Most of the coefficients for race/ethnicity remain directionally robust compared to our results using state dummies except for Black (where the direction of the effect is positive without state fixed effects) and Native American (where the direction of the effect is negative without state fixed effects). This can be explained by the fact that state fixed effects also control for geographical bias in race/ethnicity composition (specifically, there are higher Black populations in the U.S. Southeast, where heat pump penetration is also higher). Our analysis is focused on identifying racial/ethnic differences in heat pump use after controlling for these differences. While the direction of the effects of electricity and natural gas prices remains the same when removing state fixed effects, the relationship with electricity price flips (from positive to negative) when including state rather than county electricity prices. We similarly attribute this change to potential confounding between electricity prices and other state effects (e.g., policies).

To further explore the relationships between energy prices and heat pump use, we also run separate analyses for regions where electricity or other fuels (including natural gas) are dominant with (Section 8.15) and without (Section 8.16) including energy prices. In regions where electricity is the main heating fuel (assumed to be electric resistance heating), we find that there is a significant and positive relationship between electricity price and heat pump use, as in our main analysis. In regions mainly heated by other fuels, we observe the opposite (i.e., a significant and negative relationship between electricity prices and heat pump use). This reinforces our observation that there is a greater incentive to switch to a heat pump with higher electricity prices when electric resistance heating is dominant, whereas there is less incentive when other fuels (e.g., natural gas) are dominant. We also perform an additional regression on our national dataset by removing both electricity and natural gas prices. Although energy prices are an important factor influencing heat pump use, we observe no substantial differences in our sociodemographic variables compared to our national regression including energy prices. Nonetheless, the nuanced and nonlinear relationship between electricity prices and heat pump use is complex and warrants further exploration in future work.

Finally, we run analyses using more granular data at the census block group level instead of the more aggregated census tract level, for both our national (Section 8.17) and regional (Section 8.18) results. Our results with census block-group data do not show meaningful differences from our main results using census tract data, suggesting that our findings are not primarily driven by aggregation biases at this level. We also note that not all census block groups report income, resulting in approximately 50 million buildings in our census block group analysis (compared to 64 million in our main analysis). At the national level, coefficient directions remain unchanged when compared to our main analysis using census tract data with minor differences in magnitude. At the regional level, results are overall qualitatively the same as our main analysis, with a few differences. For instance, income coefficients have a larger magnitude in the net benefits region. However, the coefficient is still larger in the net costs regions, which is the same trend when using census tract data (see Figure 5 in the main paper). Other coefficients that are not significant in our main analysis, such as renter status in the neutral region, change direction from positive to negative when using census block groups. As with our main analysis, missing data at the census block group level does not correlate with other variables such as heat pump use, race/ethnicity, and renter status (with the largest coefficient equal to 0.08).

## 8.1 Basic model (race/ethnicity, climate, energy prices, and state dummies)

***Table 8.1**. Coefficient estimates and standard errors for the basic national logistic regression model for heat pump use. The marking in the coefficient estimates denotes the significance of the coefficient (i.e., *** p<0.01, ** p<0.05, * p<0.1). Variables labeled "-" are dimensionless and have been standardized (see Table 3.1 for means and standard deviations).*

| Variable | Log Odds Coefficient | S.E. |
|---|---|---|
| Black (%) | -1.551*** | 0.004 |
| Asian (%) | -4.763*** | 0.015 |
| Hispanic (%) | -1.083*** | 0.009 |
| Multiracial (%) | -3.138*** | 0.024 |
| Pacific Islanders (%) | -7.422*** | 0.165 |
| Native (%) | 1.544*** | 0.045 |
| Other (%) | -6.242*** | 0.097 |
| Heating degree-days (-) | -0.53*** | 0.005 |
| Cooling degree-days (-) | -0.317*** | 0.005 |
| Natural gas price (-) | 0.052*** | 0.001 |
| Electricity price (-) | 0.212*** | 0.002 |
| Constant | -5.005*** | 0.046 |
| State dummies | Yes | - |
| Housing type dummies | No | - |
| Pseudo $R^2$ | 0.3032 | - |
| Observations | 52,356,989 | - |

## 8.2 Basic model and income

***Table 8.2**. Coefficient estimates and standard errors for the basic national logistic regression model for heat pump use controlling for income. The marking in the coefficient estimates denotes the significance of the coefficient (i.e., *** p<0.01, ** p<0.05, * p<0.1). Variables labeled "-" are dimensionless and have been standardized (see Table 3.1 for means and standard deviations).*

| Variable | Log Odds Coefficient | S.E. |
|---|---|---|
| Black (%) | -1.533*** | 0.005 |
| Asian (%) | -4.295*** | 0.016 |
| Hispanic (%) | -1.172*** | 0.01 |
| Multiracial (%) | -3.058*** | 0.024 |
| Pacific Islanders (%) | -8.03*** | 0.168 |
| Native (%) | 1.594*** | 0.046 |
| Other (%) | -6.234*** | 0.1 |
| Heating degree-days (-) | -0.492*** | 0.005 |
| Cooling degree-days (-) | -0.291*** | 0.005 |
| Natural gas price (-) | 0.038*** | 0.001 |
| Electricity price (-) | 0.213*** | 0.002 |
| Log Income | 7.736*** | 0.072 |
| $(\text{Log Income})^2$ | -0.348*** | 0.003 |
| Constant | -48.08*** | 0.411 |
| State dummies | Yes | - |
| Housing type dummies | No | - |
| Pseudo $R^2$ | 0.3040 | - |
| Observations | 52,285,121 | - |

## 8.3 Basic model and building age

***Table 8.3***. *Coefficient estimates and standard errors for the basic national logistic regression model for heat pump use controlling for building age. The marking in the coefficient estimates denotes the significance of the coefficient (i.e., *** p<0.01, ** p<0.05, * p<0.1). Variables labeled "-" are dimensionless and have been standardized (see Table 3.1 for means and standard deviations).*

| Variable | Log Odds Coefficient | S.E. |
|---|---|---|
| Black (%) | -1.278*** | 0.005 |
| Asian (%) | -5.826*** | 0.016 |
| Hispanic (%) | -0.736*** | 0.01 |
| Multiracial (%) | -3.364*** | 0.025 |
| Pacific Islanders (%) | -6.878*** | 0.169 |
| Native (%) | 2.187*** | 0.044 |
| Other (%) | -6.421*** | 0.098 |
| Heating degree-days (-) | -0.615*** | 0.006 |
| Cooling degree-days (-) | -0.349*** | 0.005 |
| Natural gas price (-) | 0.002 | 0.001 |
| Electricity price (-) | 0.228*** | 0.002 |
| Building age (-) | -0.024*** | 0 |
| Constant | -4.128*** | 0.046 |
| State dummies | Yes | - |
| Housing type dummies | No | - |
| Pseudo $R^2$ | 0.3313 | - |
| Observations | 51,334,229 | - |

## 8.4 Basic model and tenancy status

***Table 8.4**. Coefficient estimates and standard errors for the basic national logistic regression model for heat pump use controlling for tenancy status. The marking in the coefficient estimates denotes the significance of the coefficient (i.e., *** p<0.01, ** p<0.05, * p<0.1). Variables labeled "-" are dimensionless and have been standardized (see Table 3.1 for means and standard deviations).*

| Variable | Log Odds Coefficient | S.E. |
|---|---|---|
| Black (%) | -1.343*** | 0.005 |
| Asian (%) | -4.614*** | 0.015 |
| Hispanic (%) | -0.767*** | 0.009 |
| Multiracial (%) | -2.727*** | 0.024 |
| Pacific Islanders (%) | -6.373*** | 0.165 |
| Native (%) | 1.576*** | 0.045 |
| Other (%) | -6.031*** | 0.095 |
| Heating degree-days (-) | -0.549*** | 0.005 |
| Cooling degree-days (-) | -0.312*** | 0.005 |
| Natural gas price (-) | 0.043*** | 0.001 |
| Electricity price (-) | 0.204*** | 0.002 |
| Rented (%) | -0.562*** | 0.005 |
| Vacant (%) | 0.795*** | 0.011 |
| Constant | -4.959*** | 0.046 |
| State dummies | Yes | - |
| Housing type dummies | No | - |
| Pseudo $R^2$ | 0.3040 | - |
| Observations | 52,356,193 | - |

## 8.5 Basic model and home value

***Table 8.5**. Coefficient estimates and standard errors for the basic national logistic regression model for heat pump use controlling for home value. The marking in the coefficient estimates denotes the significance of the coefficient (i.e., *** p<0.01, ** p<0.05, * p<0.1). Variables labeled "-" are dimensionless and have been standardized (see Table 3.1 for means and standard deviations).*

| Variable | Log Odds Coefficient | S.E. |
|---|---|---|
| Black (%) | -1.507*** | 0.004 |
| Asian (%) | -4.933*** | 0.016 |
| Hispanic (%) | -1.061*** | 0.009 |
| Multiracial (%) | -3.134*** | 0.024 |
| Pacific Islanders (%) | -7.206*** | 0.165 |
| Native (%) | 1.620*** | 0.045 |
| Other (%) | -6.360*** | 0.097 |
| Heating degree-days (-) | -0.517*** | 0.005 |
| Cooling degree-days (-) | -0.327*** | 0.005 |
| Natural gas price (-) | 0.043*** | 0.001 |
| Electricity price (-) | 0.212*** | 0.002 |
| Log Home value | 0.045*** | 0.001 |
| Constant | -5.587*** | 0.046 |
| State dummies | Yes | - |
| Housing type dummies | No | - |
| Pseudo $R^2$ | 0.3025 | - |
| Observations | 52,145,047 | - |

## 8.6 Basic model and housing type

***Table 8.6***. *Coefficient estimates and standard errors for the basic national logistic regression model for heat pump use controlling for housing type. The marking in the coefficient estimates denotes the significance of the coefficient (i.e., *** p<0.01, ** p<0.05, * p<0.1). Variables labeled "-" are dimensionless and have been standardized (see Table 3.1 for means and standard deviations).*

| Variable | Log Odds Coefficient | S.E. |
|---|---|---|
| Black (%) | -1.565*** | 0.004 |
| Asian (%) | -4.950*** | 0.015 |
| Hispanic (%) | -1.144*** | 0.009 |
| Multiracial (%) | -3.145*** | 0.024 |
| Pacific Islanders (%) | -7.276*** | 0.165 |
| Native (%) | 1.430*** | 0.046 |
| Other (%) | -6.271*** | 0.097 |
| Heating degree-days (-) | -0.544*** | 0.005 |
| Cooling degree-days (-) | -0.329*** | 0.005 |
| Natural gas price (-) | 0.049*** | 0.001 |
| Electricity price (-) | 0.212*** | 0.002 |
| Constant | -5.054*** | 0.046 |
| State dummies | Yes | - |
| Housing type dummies | Yes | - |
| Pseudo $R^2$ | 0.3042 | - |
| Observations | 52,356,989 | - |

## 8.7 Complete model with building age as categorical variable

***Table 8.7***. *Coefficient estimates and standard errors for the complete national logistic regression model for heat pump use replacing building age with categorical variables. The marking in the coefficient estimates denotes the significance of the coefficient (i.e., *** p<0.01, ** p<0.05, * p<0.1). Variables labeled "-" are dimensionless and have been standardized (see Table 3.1 for means and standard deviations).*

| Variable | Log Odds Coefficient | S.E. |
|---|---|---|
| Black (%) | -1.401*** | 0.005 |
| Asian (%) | -5.219*** | 0.018 |
| Hispanic (%) | -0.917*** | 0.01 |
| Multiracial (%) | -2.976*** | 0.025 |
| Pacific Islanders (%) | -8.278*** | 0.171 |
| Native (%) | 1.239*** | 0.046 |
| Other (%) | -5.785*** | 0.097 |
| Heating degree-days (-) | -0.513*** | 0.005 |
| Cooling degree-days (-) | -0.162*** | 0.005 |
| Natural gas price (-) | -0.032*** | 0.001 |
| Electricity price (-) | 0.189*** | 0.002 |
| Log Home value | 0.056*** | 0.001 |
| Rented (%) | -0.458*** | 0.006 |
| Vacant (%) | 0.422*** | 0.012 |
| Buildings pre 1940 (%) | -1.652*** | 0.011 |
| Buildings 1940s (%) | -2.423*** | 0.021 |
| Buildings 1950s (%) | -2.319*** | 0.014 |
| Buildings 1960s (%) | -2.371*** | 0.014 |
| Buildings 1970s (%) | -0.166*** | 0.011 |
| Buildings 1980s (%) | 1.244*** | 0.01 |
| Buildings 1990s (%) | -0.069*** | 0.01 |
| Buildings 2000s (%) | 0.193*** | 0.012 |
| Log Income | 1.979*** | 0.069 |
| $(\text{Log Income})^2$ | -0.117*** | 0.003 |
| Constant | -12.76*** | 0.392 |
| State dummies | Yes | - |
| Housing type dummies | No | - |
| Pseudo $R^2$ | 0.3251 | - |
| Observations | 52,076,215 | - |

## 8.8 Natural gas access

***Table 8.8**. Coefficient estimates and standard errors for the complete national logistic regression model for heat pump use replacing natural gas prices with a binary variable denoting natural gas access. The marking in the coefficient estimates denotes the significance of the coefficient (i.e., *** p<0.01, ** p<0.05, * p<0.1). Variables labeled "-" are dimensionless and have been standardized (see Table 3.1 for means and standard deviations).*

| Variable | Log Odds Coefficient | S.E. |
|---|---|---|
| Black (%) | -1.595*** | 0.005 |
| Asian (%) | -4.48*** | 0.016 |
| Hispanic (%) | -0.916*** | 0.009 |
| Multiracial (%) | -2.941*** | 0.021 |
| Pacific Islanders (%) | -8.277*** | 0.151 |
| Native (%) | 2.407*** | 0.024 |
| Other (%) | -4.558*** | 0.081 |
| Heating degree-days (-) | -0.902*** | 0.005 |
| Cooling degree-days (-) | -0.391*** | 0.004 |
| Natural gas access | -0.111*** | 0.002 |
| Electricity price (-) | 0.234*** | 0.002 |
| Log Home value | 0.034*** | 0.001 |
| Rented (%) | -0.531*** | 0.006 |
| Vacant (%) | 0.065*** | 0.008 |
| Building age (-) | -0.733*** | 0.001 |
| Log Income | 3.127*** | 0.061 |
| $(\text{Log Income})^2$ | -0.167*** | 0.003 |
| Constant | -20.056*** | 0.368 |
| State dummies | Yes | - |
| Housing type dummies | Yes | - |
| Pseudo $R^2$ | 0.3295 | - |
| Observations | 62,138,170 | - |

## 8.9 Race percent interaction with income

***Table 8.9**. Coefficient estimates and standard errors for the complete national logistic regression model for heat pump use and interactions of race percent with income. The marking in the coefficient estimates denotes the significance of the coefficient (i.e., *** p<0.01, ** p<0.05, * p<0.1). Variables labeled "-" are dimensionless and have been standardized (see Table 3.1 for means and standard deviations). An asterisk, "*", between two variables denotes an interaction, for example, "Hispanic * rented".*

| Variable | Log Odds Coefficient | S.E. |
|---|---|---|
| Black (%) | -12.164*** | 0.110 |
| Asian (%) | -3.189*** | 0.413 |
| Hispanic (%) | 18.327*** | 0.221 |
| Multiracial (%) | 3.839*** | 0.595 |
| Pacific Islanders (%) | -32.039*** | 5.197 |
| Native (%) | 15.07*** | 1.218 |
| Other (%) | 14.515*** | 2.548 |
| Black * log income | 0.974*** | 0.010 |
| Asian * log income | -0.084** | 0.036 |
| Hispanic * log income | -1.76*** | 0.020 |
| Multiracial * log income | -0.6*** | 0.053 |
| Pacific Islanders * log income | 2.131*** | 0.470 |
| Native * log income | -1.304*** | 0.114 |
| Other * log income | -1.805*** | 0.228 |
| Heating degree-days (-) | -0.495*** | 0.005 |
| Cooling degree-days (-) | -0.149*** | 0.005 |
| Natural gas price (-) | -0.014*** | 0.001 |
| Electricity price (-) | 0.212*** | 0.002 |
| Log Home value | 0.017*** | 0.001 |
| Rented (%) | -0.395*** | 0.007 |
| Vacant (%) | 0.985*** | 0.011 |
| Building age (-) | -0.733*** | 0.001 |
| Log Income | 2.331*** | 0.083 |
| (Log Income)$^2$ | -0.126*** | 0.004 |
| Constant | -15.71*** | 0.473 |
| State dummies | Yes | - |
| Housing type dummies | Yes | - |
| Pseudo $R^2$ | 0.3343 | - |
| Observations | 52,076,215 | - |

## 8.10 Race percent interaction with tenancy status

***Table 8.10**. Coefficient estimates and standard errors for the complete national logistic regression model for heat pump use and interactions of race percent with tenancy status. The marking in the coefficient estimates denotes the significance of the coefficient (i.e., *** p<0.01, ** p<0.05, * p<0.1). Variables labeled "-" are dimensionless and have been standardized (see Table 3.1 for means and standard deviations). An asterisk, "*", between two variables denotes an interaction, for example, "Hispanic * rented".*

| Variable | Log Odds Coefficient | S.E. |
|---|---|---|
| Black (%) | -1.282*** | 0.01 |
| Asian (%) | -4.556*** | 0.028 |
| Hispanic (%) | -1.047*** | 0.022 |
| Multiracial (%) | -2.355*** | 0.047 |
| Pacific Islanders (%) | -13.377*** | 0.414 |
| Native (%) | 1.707*** | 0.133 |
| Other (%) | -4.516*** | 0.193 |
| Black * rented | 1.051*** | 0.023 |
| Asian * rented | 0.522*** | 0.082 |
| Hispanic * rented | -0.249*** | 0.05 |
| Multiracial * rented | -0.441*** | 0.131 |
| Pacific Islanders * rented | 9.291*** | 0.832 |
| Native * rented | -1.353*** | 0.364 |
| Other * rented | -0.283 | 0.553 |
| Black * vacant | -12.399*** | 0.132 |
| Asian * vacant | 6.312*** | 0.495 |
| Hispanic * vacant | 5.201*** | 0.135 |
| Multiracial * vacant | -8.094*** | 0.281 |
| Pacific Islanders * vacant | 45.291*** | 2.72 |
| Native * vacant | -0.274 | 0.454 |
| Other * vacant | -25.853*** | 1.341 |
| Heating degree-days (-) | -0.516*** | 0.005 |
| Cooling degree-days (-) | -0.153*** | 0.005 |
| Natural gas price (-) | -0.02*** | 0.001 |
| Electricity price (-) | 0.215*** | 0.002 |
| Log Home value | 0.015*** | 0.001 |
| Rented (%) | -0.608*** | 0.011 |
| Vacant (%) | 1.313*** | 0.016 |
| Building age (-) | -0.735*** | 0.001 |
| Log Income | 2.433*** | 0.077 |
| $(\text{Log Income})^2$ | -0.133*** | 0.003 |
| Constant | -15.877*** | 0.437 |
| State dummies | Yes | - |
| Housing type dummies | Yes | - |
| Pseudo $R^2$ | 0.3342 | - |
| Observations | 51,139,213 | - |

## 8.11 Race percent interaction with building age

***Table 8.11**. Coefficient estimates and standard errors for the complete national logistic regression model for heat pump use and interactions of race percent with building age. The marking in the coefficient estimates denotes the significance of the coefficient (i.e., *** p<0.01, ** p<0.05, * p<0.1). Variables labeled "-" are dimensionless and have been standardized (see Table 3.1 for means and standard deviations). An asterisk, "*", between two variables denotes an interaction, for example, "Hispanic * rented".*

| Variable | Log Odds Coefficient | S.E. |
|---|---|---|
| Black (%) | -1.487*** | 0.005 |
| Asian (%) | -3.169*** | 0.018 |
| Hispanic (%) | -1.108*** | 0.011 |
| Multiracial (%) | -2.87*** | 0.028 |
| Pacific Islanders (%) | -7.279*** | 0.177 |
| Native (%) | 0.464*** | 0.068 |
| Other (%) | -6.583*** | 0.12 |
| Black * building age | -0.273*** | 0.006 |
| Asian * building age | 2.341*** | 0.016 |
| Hispanic * building age | -0.458*** | 0.013 |
| Multiracial * building age | -0.119*** | 0.033 |
| Pacific Islanders * building age | 2.686*** | 0.207 |
| Native * building age | -1.842*** | 0.08 |
| Other * building age | -1.873*** | 0.135 |
| Heating degree-days (-) | -0.497*** | 0.005 |
| Cooling degree-days (-) | -0.155*** | 0.005 |
| Natural gas price (-) | -0.016*** | 0.001 |
| Electricity price (-) | 0.213*** | 0.002 |
| Log Home value | 0.014*** | 0.001 |
| Rented (%) | -0.434*** | 0.007 |
| Vacant (%) | 0.993*** | 0.011 |
| Building age (-) | -0.737*** | 0.002 |
| Log Income | 2.952*** | 0.071 |
| $(\text{Log Income})^2$ | -0.155*** | 0.003 |
| Constant | -19.052*** | 0.405 |
| State dummies | Yes | - |
| Housing type dummies | Yes | - |
| Pseudo $R^2$ | 0.3343 | - |
| Observations | 51,139,213 | - |

## 8.12 Race percent interaction with income, tenancy, and building age

***Table 8.12**. Coefficient estimates and standard errors for the complete national logistic regression model for heat pump use and interactions of race percent with income, tenancy status, and building age. The marking in the coefficient estimates denotes the significance of the coefficient (i.e., *** p<0.01, ** p<0.05, * p<0.1). Variables labeled "-" are dimensionless and have been standardized (see Table 3.1 for means and standard deviations). An asterisk, "*", between two variables denotes an interaction, for example, "Hispanic * rented".*

| Variable | Log Odds Coefficient | S.E. |
|---|---|---|
| Black (%) | -21.37*** | 0.171 |
| Asian (%) | -17.50*** | 0.508 |
| Hispanic (%) | 26.74*** | 0.268 |
| Multiracial (%) | 17.20*** | 0.771 |
| Pacific Islanders (%) | -139.1*** | 7.331 |
| Native (%) | 24.60*** | 1.668 |
| Other (%) | 51.24*** | 2.864 |
| Black * income | 1.721*** | 0.015 |
| Asian * income | 1.232*** | 0.043 |
| Hispanic * income | -2.471*** | 0.024 |
| Multiracial * income | -1.676*** | 0.067 |
| Pacific Islanders * income | 11.02*** | 0.635 |
| Native * income | -2.146*** | 0.147 |
| Other * income | -4.878*** | 0.248 |
| Black * rented | 3.459*** | 0.030 |
| Asian * rented | 1.313*** | 0.092 |
| Hispanic * rented | -2.405*** | 0.056 |
| Multiracial * rented | -2.886*** | 0.159 |
| Pacific Islanders * rented | 19.98*** | 1.079 |
| Native * rented | -3.220*** | 0.424 |
| Other * rented | -5.220*** | 0.632 |
| Black * vacant | -4.559*** | 0.137 |
| Asian * vacant | 3.704*** | 0.532 |
| Hispanic * vacant | 1.247*** | 0.143 |
| Multiracial * vacant | -12.66*** | 0.334 |
| Pacific Islanders * vacant | 79.08*** | 3.003 |
| Native * vacant | -3.769*** | 0.528 |
| Other * vacant | -39.05*** | 1.431 |
| Black * building age | -0.108*** | 0.006 |
| Asian * building age | 2.563*** | 0.016 |
| Hispanic * building age | -0.511*** | 0.012 |
| Multiracial * building age | -0.234*** | 0.034 |
| Pacific Islanders * building age | 2.259*** | 0.211 |
| Native * building age | -1.791*** | 0.080 |
| Other * building age | -2.013*** | 0.131 |
| Heating degree-days (-) | -0.485*** | 0.005 |
| Cooling degree-days (-) | -0.133*** | 0.005 |
| Natural gas price (-) | -0.00568*** | 0.002 |
| Electricity price (-) | 0.206*** | 0.002 |
| Log Home value | 0.0148*** | 0.001 |
| Rented (%) | -0.666*** | 0.012 |
| Vacant (%) | 1.297*** | 0.016 |
| Building age (-) | -0.743*** | 0.002 |
| Log Income | 1.957*** | 0.085 |
| $(\text{Log Income})^2$ | -0.112*** | 0.004 |
| Constant | -13.22*** | 0.482 |
| State dummies | Yes | - |
| Housing type dummies | Yes | - |
| Pseudo $R^2$ | 0.3364 | - |
| Observations | 51,139,213 | - |

## 8.13 Complete model without state dummies

***Table 8.13***. *Coefficient estimates and standard errors for the national logistic regression model for heat pump use without State dummy variables. Housing type (extracted from ZTRAX) dummy variables are included in the model but not shown here for brevity. The marking in the coefficient estimates denotes the significance of the coefficient (i.e., *** p<0.01, ** p<0.05, * p<0.1). Variables labeled "-" are dimensionless.*

| Variable | Log Odds Coefficient | S.E. |
|---|---|---|
| Black (%) | 0.422*** | 0.004 |
| Asian (%) | -5.322*** | 0.015 |
| Hispanic (%) | -4.722*** | 0.010 |
| Multiracial (%) | -3.559*** | 0.022 |
| Pacific Islanders (%) | -11.283*** | 0.181 |
| Native (%) | -1.118*** | 0.052 |
| Other (%) | -7.611*** | 0.088 |
| Log Income (-) | 6.963*** | 0.066 |
| $(\text{Log Income})^2$ (-) | -0.319*** | 0.003 |
| Rented (%) | -0.307*** | 0.006 |
| Vacant (%) | -0.103*** | 0.012 |
| Log Home value (-) | 0.014*** | 0.000 |
| Building age (-) | -0.739*** | 0.001 |
| Heating degree-days (-) | -0.792*** | 0.001 |
| Cooling degree-days (-) | -0.953*** | 0.001 |
| Electricity price (-) | 0.274*** | 0.001 |
| Natural gas price (-) | 0.259*** | 0.001 |
| State dummies | No | - |
| Housing type dummies | Yes | - |
| Constant | -40.553*** | 0.370 |
| Pseudo $R^2$ | 0.125 | - |
| Observations | 50,957,182 | - |

### 8.14 Complete model without state dummies and with state level electricity prices

***Table 8.14.*** *Coefficient estimates and standard errors for the national logistic regression model for heat pump use with state level electricity prices. Housing type (extracted from ZTRAX) dummy variables are included in the model but not shown here for brevity. The marking in the coefficient estimates denotes the significance of the coefficient (i.e., *** p<0.01, ** p<0.05, * p<0.1). Variables labeled "-" are dimensionless.*

| Variable | Log Odds Coefficient | S.E. |
|---|---|---|
| Black (%) | 0.611*** | 0.004 |
| Asian (%) | -4.529*** | 0.016 |
| Hispanic (%) | -2.499*** | 0.010 |
| Multiracial (%) | -3.984*** | 0.023 |
| Pacific Islanders (%) | -23.077*** | 0.211 |
| Native (%) | -4.475*** | 0.075 |
| Other (%) | -5.632*** | 0.079 |
| Log Income (-) | 4.148*** | 0.066 |
| $(\text{Log Income})^2$ (-) | -0.172*** | 0.003 |
| Rented (%) | -0.192*** | 0.006 |
| Vacant (%) | 0.588*** | 0.012 |
| Log Home value (-) | 0.029*** | 0.001 |
| Building age (-) | -0.570*** | 0.001 |
| Heating degree-days (-) | -0.715*** | 0.002 |
| Cooling degree-days (-) | -1.320*** | 0.001 |
| Electricity price (-) | -0.918*** | 0.001 |
| Natural gas price (-) | 0.533*** | 0.001 |
| State dummies | No | - |
| Housing type dummies | Yes | - |
| Constant | -28.163*** | 0.373 |
| Pseudo $R^2$ | 0.1569 | - |
| Observations | 50,957,182 | - |

## 8.15 Complete model in electric resistance and direct fuel heating regions

***Table 8.15.*** *Coefficient estimates and standard errors for the national logistic regression model for heat pump use on regions with 1) electric resistance heating and 2) direct fuel heating (i.e., natural gas, propane, fuel oil and wood) as the main heating fuel per county (see Figure 5b in the main paper for reference). Housing type (extracted from ZTRAX) dummy variables are included in the model but not shown here for brevity. The marking in the coefficient estimates denotes the significance of the coefficient (i.e., *** p<0.01, ** p<0.05, * p<0.1). Variables labeled "-" are dimensionless.*

| Variable | Electric heating | | Direct fuel heating | |
|---|---|---|---|---|
| | Log Odds | S.E. | Log Odds | S.E. |
| Black (%) | -1.122*** | 0.007 | -0.125*** | 0.009 |
| Asian (%) | -5.414*** | 0.025 | -1.335*** | 0.022 |
| Hispanic (%) | -0.199*** | 0.012 | -0.972*** | 0.019 |
| Multiracial (%) | -2.630*** | 0.029 | -2.979*** | 0.049 |
| Pacific Islanders (%) | -10.171*** | 0.211 | 2.995*** | 0.295 |
| Native (%) | 0.299*** | 0.058 | 2.895*** | 0.080 |
| Other (%) | -5.456*** | 0.117 | -3.607*** | 0.181 |
| Log Income (-) | -0.400*** | 0.089 | 13.410*** | 0.163 |
| $(\text{Log Income})^2$ (-) | 0.005 | 0.004 | -0.596*** | 0.007 |
| Rented (%) | -0.167*** | 0.008 | -0.767*** | 0.013 |
| Vacant (%) | 1.452*** | 0.013 | 0.522*** | 0.037 |
| Log Home value (-) | 0.075*** | 0.001 | 0.073*** | 0.001 |
| Building age (-) | -0.721 *** | 0.002 | -0.573*** | 0.002 |
| Heating degree-days (-) | 0.094*** | 0.007 | -0.474*** | 0.013 |
| Cooling degree-days (-) | 0.137*** | 0.006 | -0.458*** | 0.013 |
| Electricity price (-) | 0.336*** | 0.003 | -0.259*** | 0.004 |
| Natural gas price (-) | -0.122*** | 0.002 | -0.033*** | 0.004 |
| State dummies | Yes | - | Yes | - |
| Housing type dummies | Yes | - | Yes | - |
| Constant | -3.355*** | 0.495 | -84.681*** | 0.992 |
| Pseudo $R^2$ | 0.3248 | - | 0.2200 | - |
| Observations | 19,180,512 | - | 31,831,866 | - |

## 8.16 Complete model without energy prices

***Table 8.16.*** *Coefficient estimates and standard errors for the national logistic regression model for heat pump use without electricity and natural gas prices. Housing type (extracted from ZTRAX) dummy variables are included in the model but not shown here for brevity. The marking in the coefficient estimates denotes the significance of the coefficient (i.e., *** p<0.01, ** p<0.05, * p<0.1). Variables labeled "-" are dimensionless.*

| Variable | Log Odds Coefficient | S.E. |
|---|---|---|
| Black (%) | -1.587*** | 0.004 |
| Asian (%) | -4.503*** | 0.016 |
| Hispanic (%) | -0.960*** | 0.010 |
| Multiracial (%) | -2.907*** | 0.0210 |
| Pacific Islanders (%) | -8.097*** | 0.151 |
| Native (%) | 2.517*** | 0.024 |
| Other (%) | -4.526*** | 0.080 |
| Log Income (-) | 3.094*** | 0.061 |
| $(\text{Log Income})^2$ (-) | -0.167*** | 0.003 |
| Rented (%) | -0.591*** | 0.006 |
| Vacant (%) | 0.170*** | 0.008 |
| Log Home value (-) | 0.034*** | 0.001 |
| Building age (-) | -0.735*** | 0.001 |
| Heating degree-days (-) | -0.956*** | 0.004 |
| Cooling degree-days (-) | -0.403*** | 0.004 |
| State dummies | Yes | - |
| Housing type dummies | Yes | - |
| Constant | -21.070*** | 0.342 |
| Pseudo $R^2$ | 0.3282 | - |
| Observations | 62,017,875 | - |

## 8.17 Regressions results with census block group data (national)

***Table 8.17***. *Coefficient estimates and standard errors for the national logistic regression model for heat pump use using census block group demographic data. State and housing type (extracted from ZTRAX) dummy variables are also included in the model but not shown here for brevity. The marking in the coefficient estimates denotes the significance of the coefficient (i.e., *** p<0.01, ** p<0.05, * p<0.1). Variables labeled "-" are dimensionless.*

| Variable | Log Odds Coefficient | S.E. |
|---|---|---|
| Black (%) | -1.218*** | 0.005 |
| Asian (%) | -3.701*** | 0.014 |
| Hispanic (%) | -1.017*** | 0.008 |
| Multiracial (%) | -1.649*** | 0.021 |
| Pacific Islanders (%) | -3.935*** | 0.119 |
| Native (%) | 0.835*** | 0.043 |
| Other (%) | -3.391*** | 0.067 |
| Log Income (-) | 5.107*** | 0.059 |
| $(\text{Log Income})^2$ (-) | -0.246*** | 0.003 |
| Rented (%) | -0.307*** | 0.005 |
| Vacant (%) | 0.994*** | 0.01 |
| Log Home value (-) | 0.010*** | 0.001 |
| Building age (-) | -0.745*** | 0.001 |
| Heating degree-days (-) | -0.479*** | 0.006 |
| Cooling degree-days (-) | -0.213*** | 0.005 |
| Electricity price (-) | 0.217*** | 0.002 |
| Natural gas price (-) | -0.010*** | 0.001 |
| State dummies | Yes | - |
| Housing type dummies | Yes | - |
| Constant | -31.783*** | 0.337 |
| Pseudo $R^2$ | 0.329 | - |
| Observations | 49,613,213 | - |

## 8.18 Regressions results with census block group data (regional)

***Table 8.18***. *Coefficient estimates and standard errors for the regional logistic regression models for heat pump use using census block group demographic data, where regions are defined based on household benefits (in the form of reduced energy bills). State and housing type (extracted from ZTRAX) dummy variables are also included in the model but not shown here for*

*brevity. The marking in the coefficient estimates denotes the significance of the coefficient (i.e., *** p<0.01, ** p<0.05, * p<0.1). Variables labeled "-" are dimensionless.*

| Variable | Net benefits | | Neutral | | Net costs | |
|---|---|---|---|---|---|---|
| | **Log Odds** | **S.E.** | **Log Odds** | **S.E.** | **Log Odds** | **S.E.** |
| Black (%) | -1.218*** | 0.005 | -2.184*** | 0.031 | -0.315*** | 0.021 |
| Asian (%) | -3.701*** | 0.018 | -2.346*** | 0.053 | -0.828*** | 0.024 |
| Hispanic (%) | -1.017*** | 0.009 | -0.929*** | 0.029 | -0.359*** | 0.021 |
| Multiracial (%) | -1.649*** | 0.024 | -2.336*** | 0.077 | -0.075 | 0.068 |
| Pacific Islanders (%) | -3.935*** | 0.143 | -2.406*** | 0.283 | 10.542*** | 0.466 |
| Native (%) | 0.835*** | 0.052 | 1.598*** | 0.106 | 0.336* | 0.183 |
| Other (%) | -3.391*** | 0.077 | -0.733*** | 0.209 | -1.664*** | 0.196 |
| Log Income (-) | 5.107*** | 0.065 | 3.498*** | 0.224 | 9.626*** | 0.221 |
| $(\text{Log Income})^2$ (-) | -0.246*** | 0.003 | -0.165*** | 0.01 | -0.434*** | 0.01 |
| Rented (%) | -0.307*** | 0.006 | -0.127*** | 0.018 | -1.097*** | 0.018 |
| Vacant (%) | 0.994*** | 0.011 | 1.789*** | 0.034 | -0.575*** | 0.067 |
| Log Home value (-) | 0.01*** | 0.001 | 0.156*** | 0.002 | 0.075*** | 0.002 |
| Building age (-) | -0.745*** | 0.001 | -0.594*** | 0.004 | -0.154*** | 0.003 |
| Heating degree-days (-) | -0.479*** | 0.006 | 0.158*** | 0.019 | -1.598*** | 0.031 |
| Cooling degree-days (-) | -0.213*** | 0.005 | 0.748*** | 0.018 | -2.003*** | 0.029 |
| Electricity price (-) | 0.217*** | 0.003 | 0.028*** | 0.008 | -0.9*** | 0.012 |
| Natural gas price (-) | -0.01*** | 0.002 | -0.012 | 0.008 | -0.064*** | 0.009 |
| State dummies | Yes | - | Yes | - | Yes | - |
| Housing type dummies | Yes | - | Yes | - | Yes | - |
| Constant | -19.994*** | 0.364 | -28.416 | 1.265 | -58.939 | 1.298 |
| Pseudo $R^2$ | 0.3145 | - | 0.1935 | - | 0.2607 | - |
| Observations | 26,347,061 | - | 14,015,954 | - | 9,248,971 | - |

## Section 9: Supplementary Figures

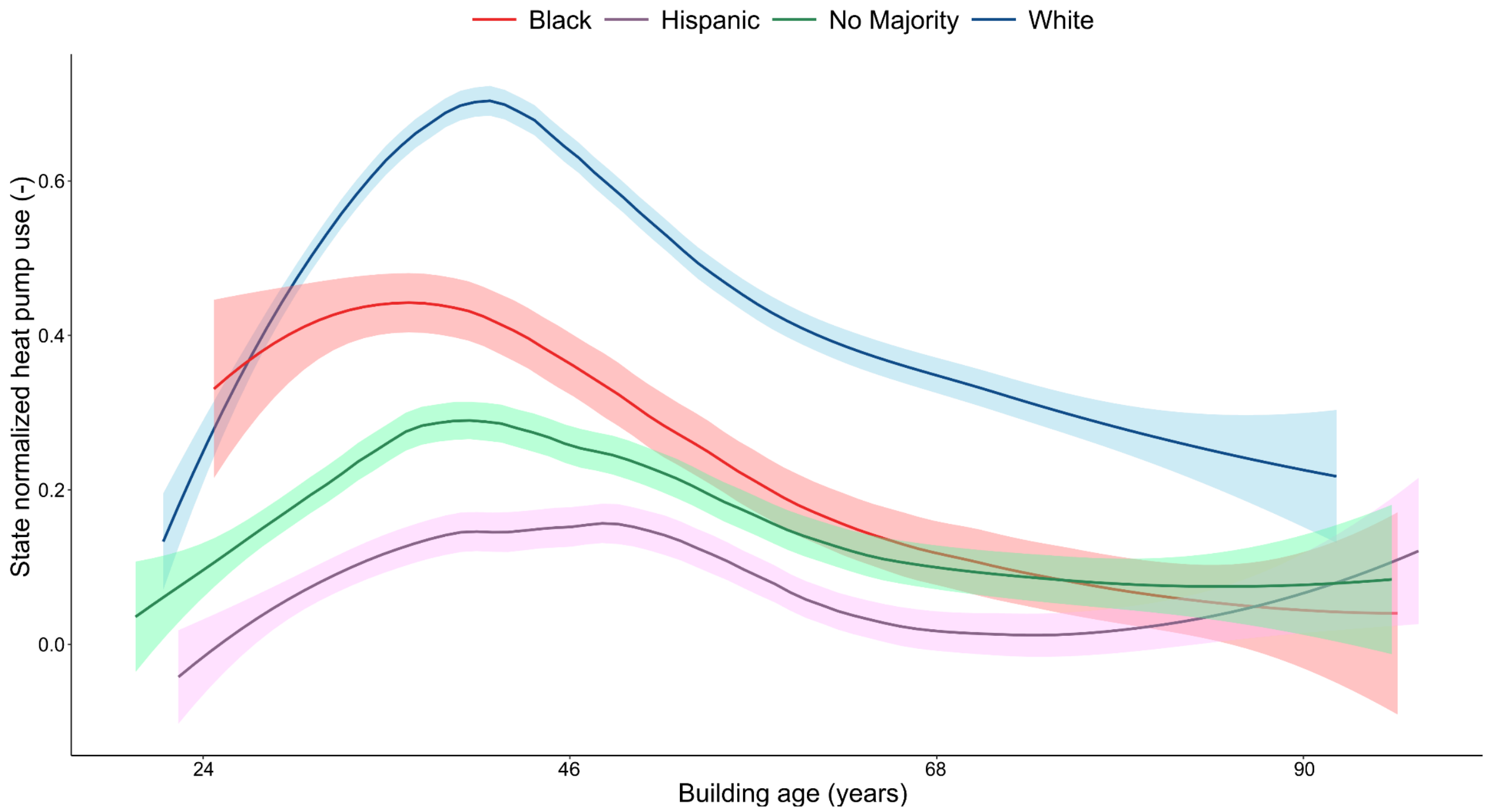

***Figure 9.1.*** *Relationship between building age and heat pump use using a nonparametric analysis (see Figure 2 in the main paper for building age relationship with heat pump use using a parametric analysis). Lines show LOESS fits and shaded regions show 90% confidence intervals. For better visualization, the building age has been restricted to the middle 98 percentile.*

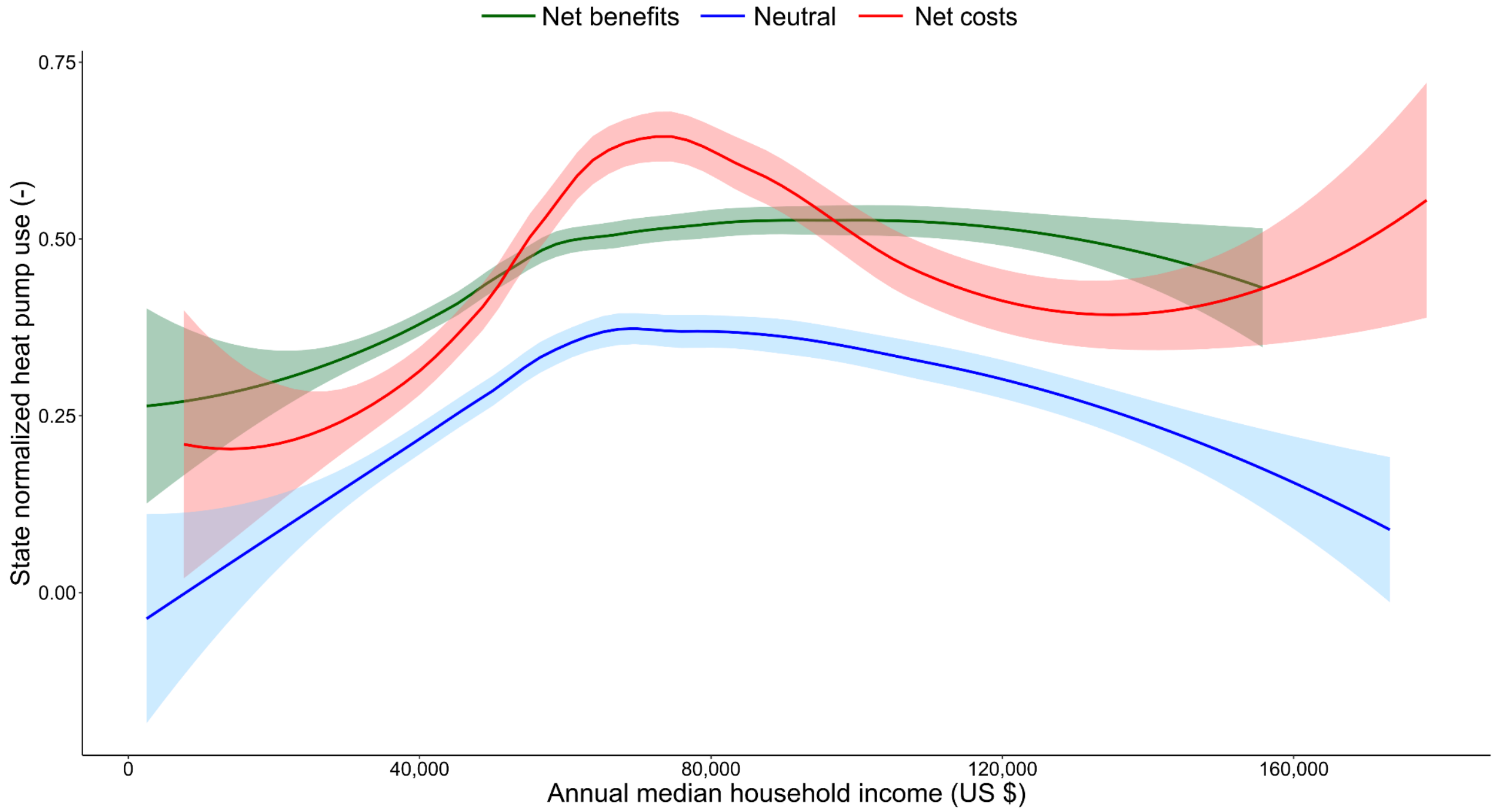

***Figure 9.2.*** *Relationship between income and heat pump use for each heat pump cost/benefit region using a nonparametric analysis (see Figure 5 in the main paper for comparison). Lines show LOESS fits and shaded regions show 90% confidence intervals.*

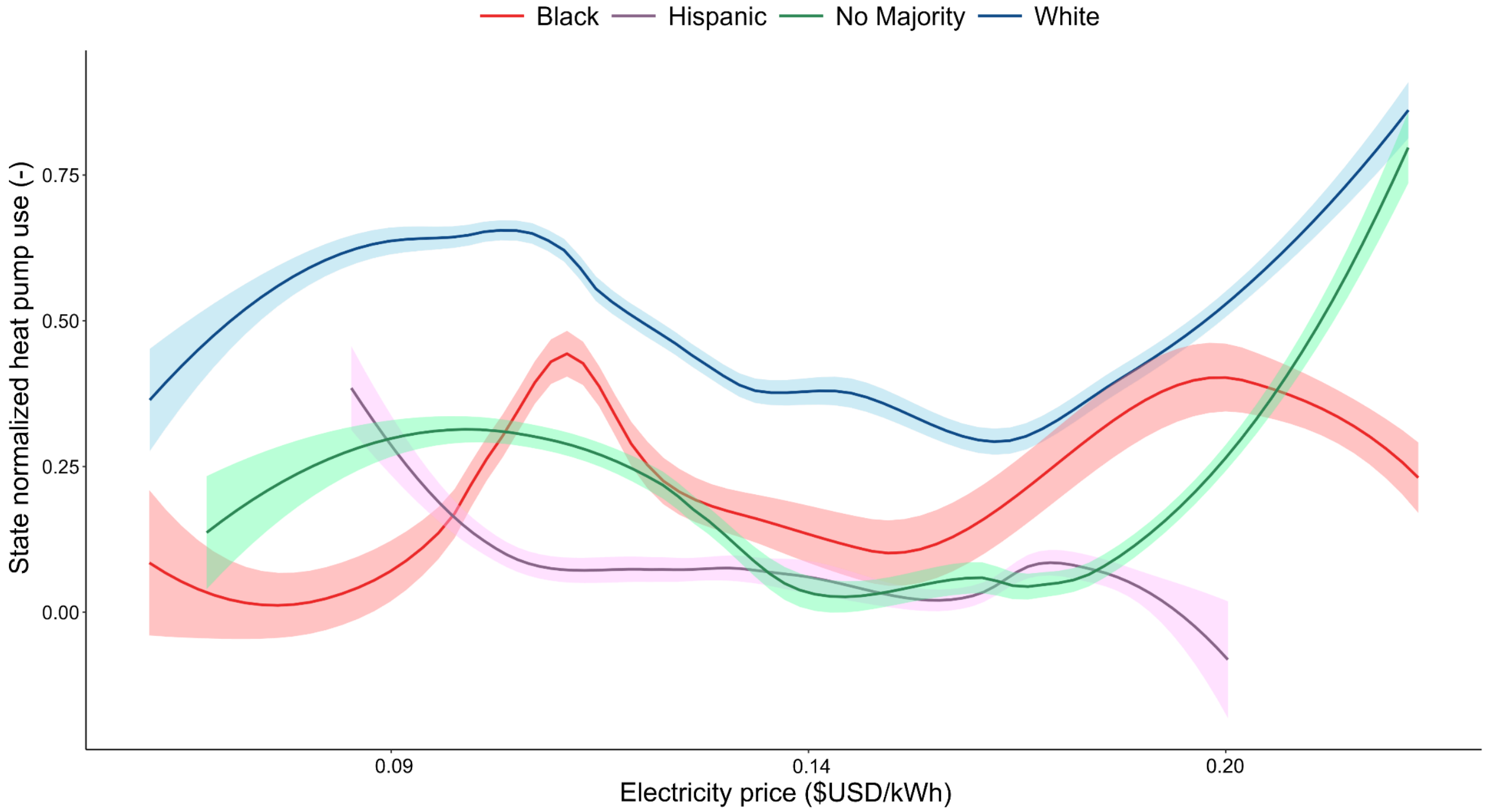

***Figure 9.3.*** *Relationship between electricity price and heat pump use using a nonparametric analysis (see Figure 3 in the main paper for electricity price relationship with heat pump use using a parametric analysis). Lines show LOESS fits and shaded regions show 90% confidence*

*intervals. For better visualization, the electricity price has been restricted to the middle 98 percentile.*